\documentclass[sigconf]{acmart}
\AtBeginDocument{%
  }
    
\usepackage[english]{babel}
\usepackage{graphicx}
\usepackage{dirtytalk}
\usepackage{amsmath,empheq}
\usepackage{mathrsfs}
\usepackage{tikz}
\usepackage{neuralnetwork}
\usepackage{subfig}
\usepackage{booktabs}
\usepackage[ruled,vlined,linesnumbered]{algorithm2e}
\usepackage{setspace}
\usepackage{xcolor}
\usepackage{url}
\usepackage{enumerate}
\usepackage{diagbox}
\usepackage{rotating}
\usepackage{float}
\usepackage{hyperref}
\usepackage{comment}
\usepackage{multirow}
\usepackage{nopageno}
\usepackage{enumitem}
\usepackage[normalem]{ulem}
\usepackage{soul}
\usepackage{balance}
\usepackage{pifont}

\usepackage{amstext}
\newcommand*{\backwardsz}{%
  \text{\reflectbox{$\mathsf{z}$}}%
}

\newtheorem{definition}{Definition}
\newtheorem{theorem}{Theorem}
\newtheorem{corollary}{Corollary}

\newtheorem{example}{Example}

\newcommand\method{LBMC}

\setcopyright{acmcopyright}
\copyrightyear{2024}
\acmYear{2024}
\acmDOI{XXXXXXX.XXXXXXX}

\acmPrice{15.00}
\acmISBN{978-1-4503-XXXX-X/18/06}

\begin{document}
\title{Efficient Cost Modeling of Space-filling Curves}

\author{Guanli Liu}
\affiliation{%
  \institution{The University of Melbourne}
  \country{Australia}
}
\email{guanli@student.unimelb.edu.au}

\author{Lars Kulik}
\affiliation{%
  \institution{The University of Melbourne}
  \country{Australia}
}
\email{lkulik@unimelb.edu.au}

\author{Christian S. Jensen}
\affiliation{%
  \institution{Aalborg University}
  \country{Denmark}
}
\email{csj@cs.aau.dk}

\author{Tianyi Li}
\affiliation{%
  \institution{Aalborg University}
  \country{Denmark}
}
\email{tianyi@cs.aau.dk}

\author{Jianzhong Qi}
\affiliation{%
  \institution{The University of Melbourne}
  \country{Australia}
}
\email{jianzhong.qi@unimelb.edu.au}

\renewcommand{\shortauthors}{\emph{et al.}}
\begin{abstract}
A \textit{space-filling curve} (SFC) maps points in a multi-dimensional space to one-dimensional points by discretizing the multi-dimensional space into cells and imposing a linear order on the cells. This way, an SFC enables the indexing of multi-dimensional data using a one-dimensional index such as a B$^+$-tree. Choosing an appropriate SFC is crucial, as different SFCs have different effects on query performance. Currently, there are two primary strategies: 1)~deterministic schemes, which are computationally efficient but often yield suboptimal query performance, and 2)~dynamic schemes, which consider a broad range of candidate SFCs based on cost functions but incur significant computational overhead. Despite these strategies, existing methods cannot efficiently measure the effectiveness of SFCs under heavy query workloads and numerous SFC options.

To address this problem, we propose means of \emph{constant-time} cost estimations that can enhance existing SFC selection algorithms, enabling them to learn more effective SFCs. Additionally, we propose an SFC learning method that leverages reinforcement learning and our cost estimation to choose an SFC pattern efficiently. Experimental studies offer evidence of the effectiveness and efficiency of the proposed means of cost estimation and SFC learning.
\end{abstract}


\maketitle

\section{Introduction}~\label{sec:intro}
\begin{figure}[h]
\vspace{-3mm}
\centering
 	\subfloat[Curve 1 (ZC)~\label{fig:ZC_XYXYXY}]{
				\includegraphics[width=0.15\textwidth]{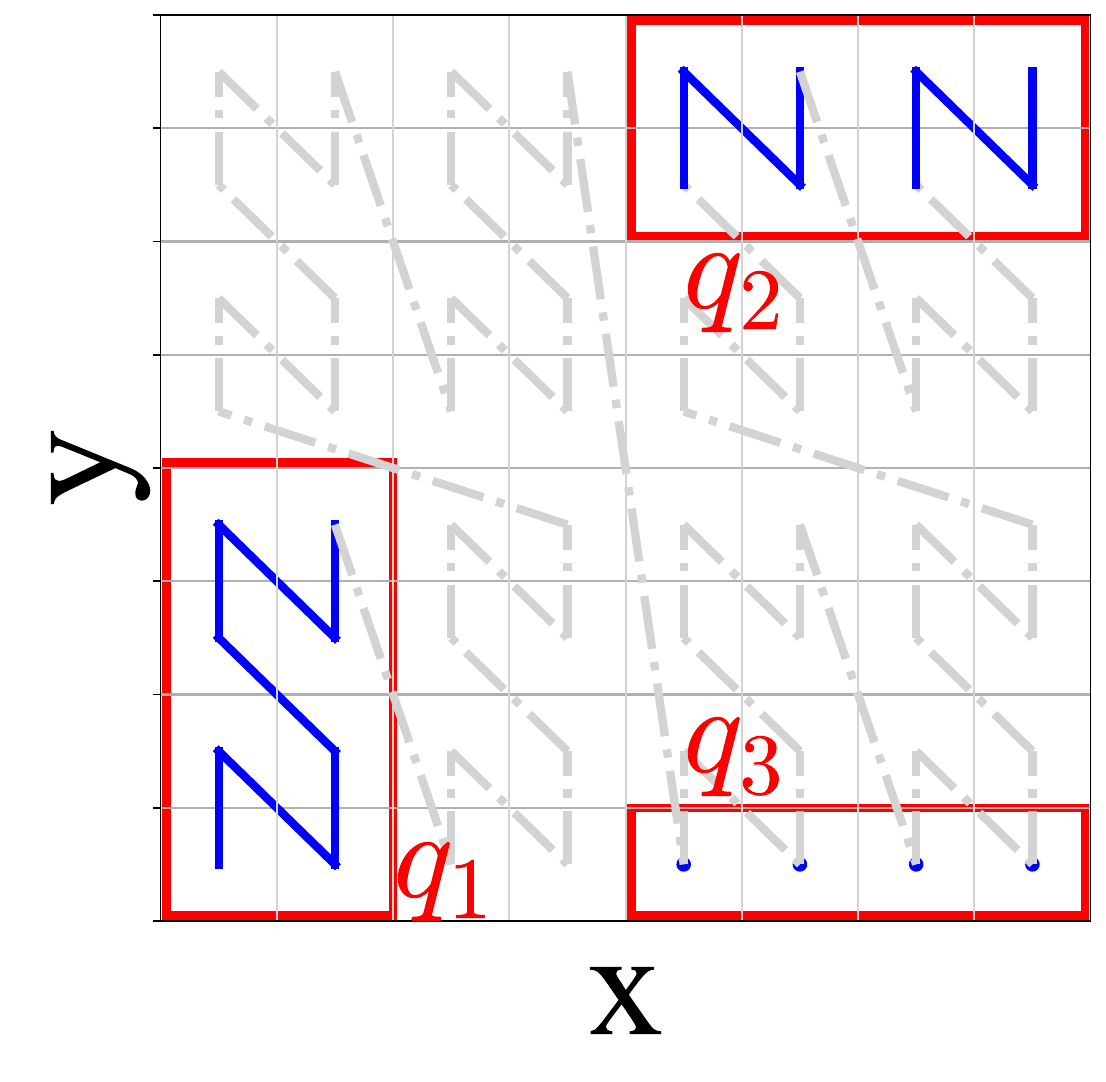}
	}
 	\subfloat[Curve 2 (ZC)~\label{fig:ZC_YXYXYX}]{
				\includegraphics[width=0.15\textwidth]{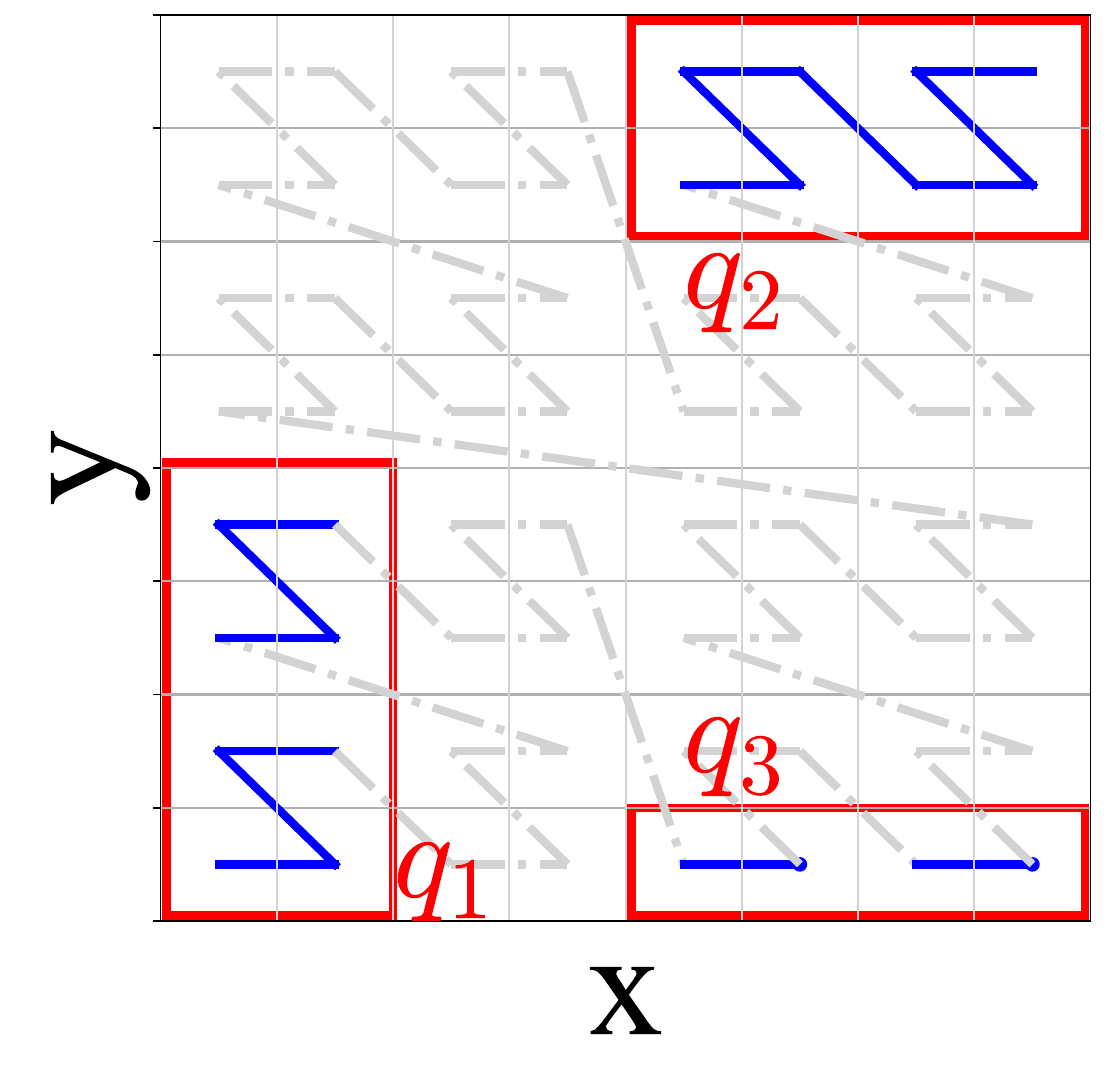}
	}
        \subfloat[Curve 3 (LC)~\label{fig:ZC_YYYXXX}]{
				\includegraphics[width=0.15\textwidth]{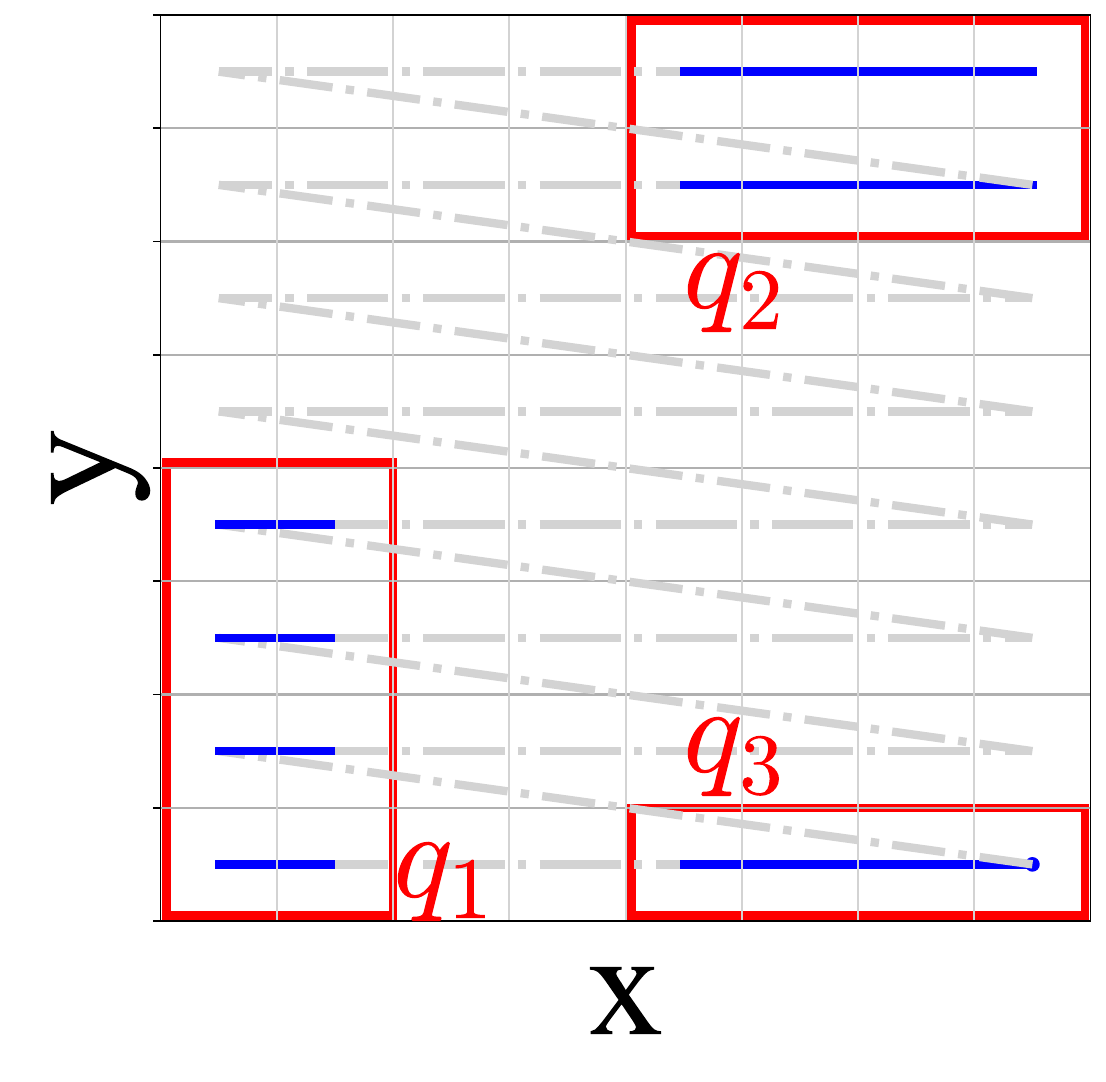}
	}
	\caption{Examples of SFCs (in grey) and queries (in red).}\label{fig:sfcs}
\end{figure}

Indexing is essential to enable efficient query processing on increasingly massive data, including spatial and other low-dimensional data. In this setting, indices based on \textit{space-filling curves} (SFC) are used widely. For example, \textit{Z-order curves} (ZC, see~Figures~\ref{fig:ZC_XYXYXY} and~\ref{fig:ZC_YXYXYX})~\cite{Zcurve} are used in 
Hudi~\cite{Hudi}, RedShift~\cite{RedShift}, and SparkSQL~\cite{SparkSQL}; \textit{lexicographic-order curves}~(LC, see Figure~\ref{fig:ZC_YYYXXX}) are used in PostgreSQL~\cite{postgresql} and SQL Server~\cite{SQLServer}; and \textit{Hilbert curves}~(HC)~\cite{Hcurve} are used in Google S2~\cite{S2}. Next, the arguably most important type of query in this setting is the range query that also serves as a foundation for other queries, including $k$NN queries.

The most efficient query processing occurs when the data needed for a query result is stored consecutively, or when the data is stored in a few data blocks. Thus, the storage organization---the order in which the data is stored---affects the cost of processing a query profoundly. When indexing data using SFC-based indices, the choice of which SFC to use for ordering the data is important.

Different range queries benefit differently from different SFCs. In Figure~\ref{fig:sfcs}, three SFCs on the
same data space are shown along with three queries. The fewer disconnected segments of an SFC that need to be accessed to compute a query, the better. To compute $q_1$, the SFC in Figure~\ref{fig:ZC_XYXYXY} is preferable because only a single segment needs to be accessed. Put differently, the data needed may be in a single or in consecutive blocks. In contrast, the SFCs in Figures~\ref{fig:ZC_YXYXYX} and~\ref{fig:ZC_YYYXXX} map the needed data to two and four segments, respectively.

Next, we observe that no single SFC is optimal for all queries. While the SFC in Figure~\ref{fig:ZC_XYXYXY} is good for $q_1$, it is suboptimal for $q_2$ and $q_3$. It is thus critical to select the right SFC for a given query (or query workload). This in turn calls for efficient means of estimating the cost of computing a query using a particular SFC (without query execution) to guide SFC selection.

Existing studies~\cite{HC_theory, SFC_theoretical_result} provide \textit{cost estimations} based on counting the number of clusters (continuous curve segments) covered by a query. However, their calculations rely on curve segment scans that require $O(V)$ time, where $V$ is proportional to the size of a query. Given a workload of $n$ queries and $m$ candidate SFCs, $O(n\cdot m\cdot V)$ time is needed to choose an SFC. This is expensive given large $n$ and $m$ (e.g., a $k\times k$ grid can form $m=k^2!$ candidate SFCs), thus jeopardizing the applicability of the cost model.

In this paper, we provide efficient means of SFC cost estimation such that a query-optimal SFC can be found efficiently. Specifically, we present algorithms that compute the cost of a query in $O(1)$ time. After an $O(n)$-time initialization, the algorithms compute the cost of $n$ queries in $O(1)$ time for each new SFC to be considered. This means that given $m$ candidate SFCs, our algorithms can find the optimal SFC in $O(m)$ time, which is much smaller than $O(n\cdot m\cdot V)$ and thus renders SFC cost estimation practical.

Our algorithms are based on a well-chosen family of SFCs, the \textit{bit-merging curves} (BMC)~\cite{QUILTS,LMSFC}. A BMC maps multi-dimensional points by merging the bit sequences of the point coordinates (i.e., column indices) from all $d$ dimensions (detailed in Section~\ref{sec:bit_merging}). We consider BMCs for two reasons: (1) BMCs generalize ZC and LC used in real systems~\cite{postgresql,Hudi,SparkSQL,SQLServer}. Algorithms to find optimal BMCs can be integrated seamlessly into real systems. (2)~The space of BMCs is large. For example, in a 2-dimensional space ($d=2$), where each dimension uses 16 bits ($\ell=16$) for a column index, there are $k = 2^\ell$ columns in each dimension of the grid. This yields about $6\times10^{8}$ (i.e., $\frac{(d \cdot  \ell)!}{(\ell!)^d}$) candidate BMCs. An efficient cost model enables finding a query-efficient SFC in this large space.

Our algorithms model the cost of a range query based on the number and lengths of curve segments covered by the query, which in turn relate to the difference between the curve values of the
end points of each curve segment. We exploit the property that the curve values of a BMC come from merging the bits of the column indices. This property enables deriving a closed-form equation to
compute the length of a curve segment in $O(d\cdot \ell) = O(1)$ time (given that $d$ and $\ell$ are constants) for $n$ queries. The property also enables pre-computing $d$ look-up tables that allow computing the number of curve segments in $O(d\cdot \ell) = O(1)$ time. Thus, we achieve constant-time SFC cost estimation.

We show the applicability of the cost estimation algorithms by incorporating them into the state-of-the-art learned BMC-based structure, the \emph{BMTree}~\cite{BMTree}.  
The BMTree computes empirical query costs by executing a query workload on the dataset to be indexed. Even with its dataset sampling strategy to reduce the computational costs for query cost estimation, the original SFC learning algorithm of the BMTree takes seven hours (cf.~BMTree-SP in Figure~\ref{fig:vary_datasize_time}) to index a dataset of 100 million points (with only 100,000 sampled points for query cost estimation). Our cost estimation algorithms bring this time down to 57 seconds (cf.~BMTree-GC in Figure~\ref{fig:vary_datasize_time}) with little impact on query efficiency.

Furthermore, we develop an SFC learning algorithm named \emph{\method} that uses Reinforcement Learning (RL) techniques to find the optimal BMC. 
Importantly, the reward calculation in RL leverages our closed-form cost estimation equation and pre-computed look-up tables, thus making the entire learning process extremely efficient. This enables the RL agent to converge rapidly to near-optimal solutions while navigating the state space.

In summary, the paper makes the following contributions:

(1) We propose algorithms for efficient range query cost estimation when using BMC-based indices on multi-dimensional datasets. The algorithms can compute the cost of a range query in $O(1)$ time as well as the cost of a workload of $n$ queries in $O(1)$ time, after a simple scan over the queries.
(2) We generalize the applicability of the cost estimation to existing state-of-the-art SFC learning methods based on BMCs, enhancing the learning efficiency of such methods. 
(3) 
We propose \method, an efficient BMC learning algorithm that leverages the proposed cost estimation. 
(4) We evaluate the cost estimation and \method\ algorithms on both real and synthetic datasets, finding that (i) our cost estimation outperforms baselines consistently by up to $10^5$ times in efficiency, (ii) our cost estimation accelerates the reward calculation of the BMTree by 400x with little impact on query efficiency,
and (iii) the \method\ algorithm has lower learning and query costs than the competing SFC learning algorithms, including the BMTree.

The rest of the paper is organized as follows. Section~\ref{sec:related_work} covers related work.
Section~\ref{sec:preliminaries} presents preliminaries, and Section~\ref{sec:technique} details our cost estimations. 
Section~\ref{sec:learning_bmc} presents \method, and 
Section~\ref{sec:experiments} reports the experimental results. Section~\ref{sec:conclusions} concludes the paper.

\section{Related Work}~\label{sec:related_work}
\textbf{Space-filling curves.}
SFCs find use in many fields, including in indexing~\cite{HRtree,ZRtree,onion,Peano}, data mining~\cite{SFC_mining}, 
 and machine learning~\cite{SFC_cnn,SFC_cnn2}. 
An SFC maps multi-dimensional data values to one-dimensional values, 
which are then indexed using a one-dimensional index, e.g., the B$^+$-tree. 

Two popular SFCs, \textit{ZC}~\cite{Zcurve} and \textit{HC}~\cite{Hcurve}, are being deployed in practical data systems~\cite{Hudi,SparkSQL,RedShift}.
Bit-merging curves (BMCs, detailed in Section~\ref{sec:bit_merging}) are a family of SFCs, where the curve value of a grid cell is formed by merging the bits of the cell's column indices from all $d$ dimensions. 
To better order the data points for specific query workloads,
\textit{QUILTS}~\cite{QUILTS} provides a heuristic method
to design a series of BMCs and selects the optimal one.
A recent technique, the 
\emph{Bit Merging Tree} (BMTree)~\cite{BMTree}, learns piece-wise SFCs (i.e., BMCs) by using a quadtree~\cite{quadtree}-like strategy to partition the  data space and selecting different BMCs for different space partitions.

\textbf{Cost estimation for space-filling curves.}  
To learn an optimal SFC, cost estimation is employed to approximate the query costs without actually computing the queries. Two studies~\cite{SFC_theoretical_result,HC_theory} offer theoretical means of estimating the number of curve segments covered by a query range.
They do not offer empirical results or guidance on how to construct a query-efficient SFC index.

QUILTS formulates the query cost $\mathcal{C}_t$ for a BMC index over a set of queries as $\mathcal{C}_t=\mathcal{C}_g\cdot \mathcal{C}_l$, where $\mathcal{C}_g$ is a \emph{global cost} and $\mathcal{C}_l$ is a \emph{local cost}. The global cost is the length of a continuous BMC segment that is able to cover a query range $q$ fully minus the length of the BMC segments in $q$, for each query. The idea is to count the number of segments outside $q$ that may need to be visited to compute the queries. The local cost is the entropy of the relative length of each segment of the BMC curve outside $q$ counted in the global cost, which reflects how uniformly distributed the lengths of such segments are. 
However, computing these two costs relies on the accumulated length of the curve segments outside $q$, which is expensive to compute.  
Given $n$ range queries, it takes $O(n \cdot c_t )$ time to compute $\mathcal{C}_t$, where $O(c_t)$ is the average estimation cost per query. Further, they can only be used to estimate the query costs of a given BMC index and do not enable an efficient search for a query-efficient BMC index.

The BMTree estimates query costs using data points sampled from the target dataset.
Such cost estimations are expensive for large datasets and many queries. For example, BMTree curve learning over a dataset of 100 million points (with 100,000 sampled points) and 1,000 queries can take more than seven hours (cf.~BMTree-SP in Figure~\ref{fig:vary_datasize_time}).
While using a smaller sampled dataset and fewer queries may reduce the learning time, the resulting curve may cause suboptimal query performance (cf.~BMTree-SP-6/8/10 in Figure~\ref{fig:vary_depth_sample_rate}).
LMSFC~\cite{LMSFC}, another recent proposal, learns a parameterized SFC (which is effectively a BMC) using Bayesian optimization~\cite{SMBO}. Like the BMTree, LMSFC uses a sampled dataset and a query workload for query cost estimation and thus has the same issues as the BMTree. 
Our study aims to address these issues by providing a highly effective and efficient cost estimation.

\textbf{Space-filling curve-based indices.}
The Hilbert R-tree~\cite{HRtree} is a classic index structure based on SFC. It uses an HC to map and order multi-dimensional data, based on which an R-tree is built on the data. This simple structure has been shown to be competitive in many follow-up studies. A recent study further achieves worst-case optimal range query processing by adding an extra \emph{rank space}-based mapping step over the input data before the Hilbert R-tree is built~\cite{rank_space}.  Another index,  the \textit{Instance-Optimal Z-Index}~\cite{optimal_z_index}, 
uses a quadtree-like strategy to recursively partition the data space. It creates four sub-spaces of a (sub-)space, which may be of different sizes. The four sub-spaces are each ordered by ZCs of different sizes and follow a `{\Large{\backwardsz}}' or an `N' shape. At the bottom level of the space partitioning hierarchy, the ZCs of sub-spaces that come from different parent sub-spaces are connected following the order of the ZC that traverses the parent sub-spaces. This way, a curve is formed that traverses all bottom-level sub-spaces, and the data points are indexed in that order. 

In the recent wave of machine learning-based optimization for indices~\cite{RMI, Flood}, SFCs have been used to order and map multi-dimensional data points to one-dimensional values, such that one-dimensional learned indices (e.g., RMI~\cite{RMI}) can be applied. ZM~\cite{ZM} and RSMI~\cite{RSMI} are representative proposals.  As the BMTree~\cite{BMTree} work shows, different learned SFCs can be plugged into these index structures to  (possibly) improve their query performance. 
Our cost estimations can be applied to further  enhance the SFC learning process as discussed above, which are orthogonal to these studies.

\section{Preliminaries}~\label{sec:preliminaries}
We start with core concepts underlying BMCs and list frequently used symbols in
Table~\ref{tab:symbols}.

    \vspace{2mm}
 \begin{table}[h]
  \small
    \centering
    \caption{Frequently used symbols.}\label{tab:symbols}
    \setlength{\tabcolsep}{1pt}
    \begin{tabular}{c|l}
    \toprule
      Symbol& Description\\
      \midrule
      \midrule
      $d$ & The data space dimensionality\\
      \hline
      $\ell$ & The number of bits for grid cell numbering in each dimension\\
      \hline
      $D$& A multi-dimensional dataset\\
      \hline
      $p$ & A data point\\
      \hline
      $q$ & A range query\\
      \hline
      $Q$ & A set of range queries\\
      \hline
      $B$ & The block size\\
      \hline
      $p_s$, $p_e$ & The start and end points on an SFC of a range query  \\
      \hline
      $n$ & The number of range queries\\
      \hline
      $\sigma$ & A bit-merging curve (BMC)\\
      \hline
      $\mathcal{F}_\sigma$ & The curve value calculation function over BMC $\sigma$ \\
      \hline
      $\alpha_{i}^{j}$ & The $j$th bit value in dimension $i$ \\
      \hline
      $\gamma_{i}^{j}$ & The position (0-indexed) of $\alpha_{i}^{j}$ in a BMC $\sigma$\\
      \hline
      $x_i$ & A value in dimension $i$\\
      \hline
      $[x_{s, i}$, $x_{e, i}]$ & A value range in dimension $i$\\
      \bottomrule
    \end{tabular}
  \end{table}

\subsection{BMC Definition}~\label{sec:bit_merging} 
A BMC maps multi-dimensional points by merging the \textit{bit sequences} of the coordinates (i.e., column indices) from all $d$ dimensions into a single bit sequence that becomes a one-dimension value~\cite{QUILTS}. 

\begin{figure}[h]
\centering
\hspace{-2mm}
		\includegraphics[width=0.155\textwidth]{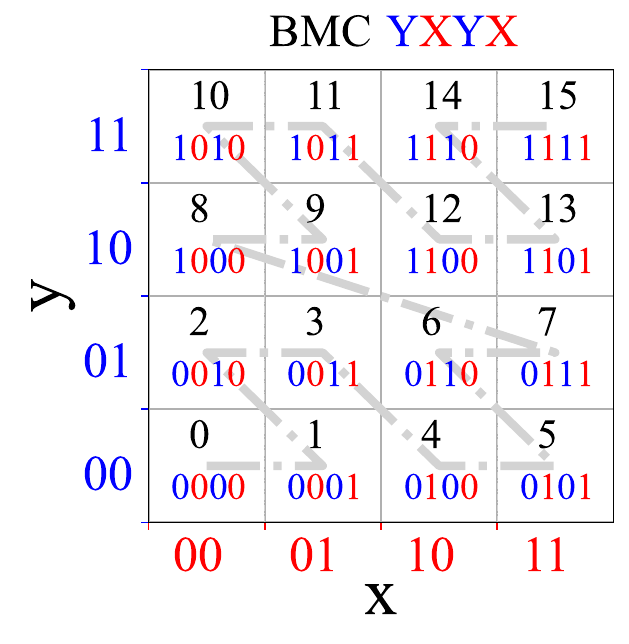}
	\hspace{0mm}
		\includegraphics[width=0.155\textwidth]{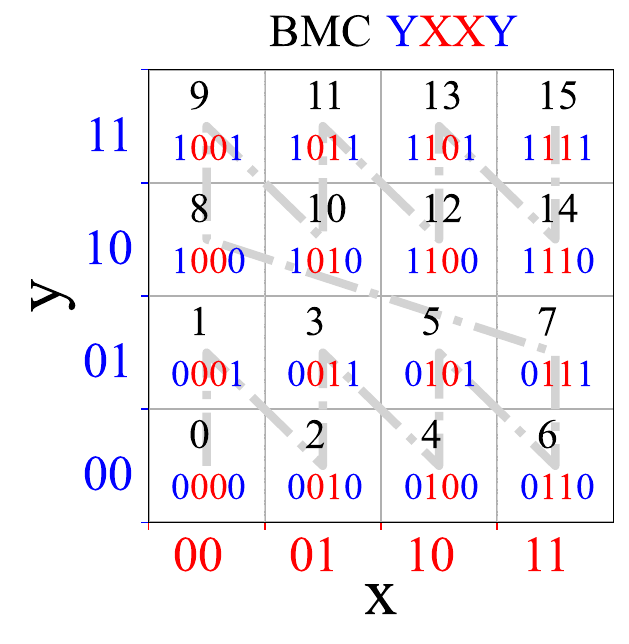}
	\hspace{0mm}
		\includegraphics[width=0.155\textwidth]{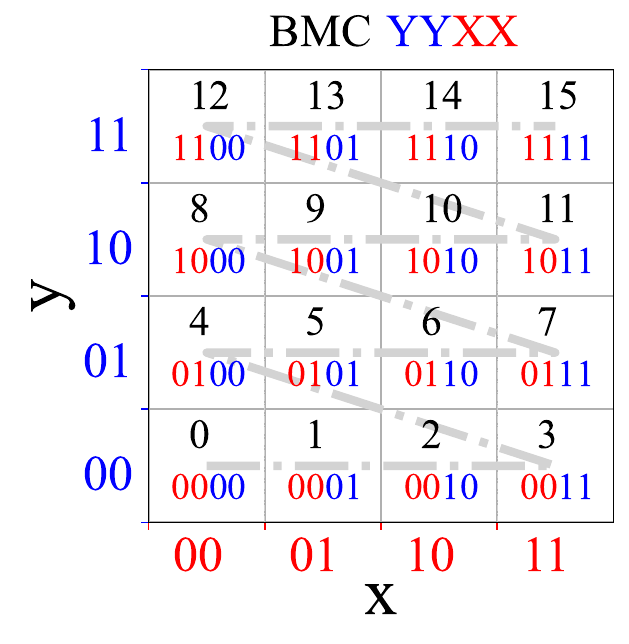}
        \vspace{-5mm}
	\caption{BMC examples ($d=2$ and $\ell=2$).}\label{fig:bit_meging_curves}
\end{figure}

Figure~\ref{fig:bit_meging_curves} plots three BMC schemes, which are represented by YXYX, YXXY, and YYXX. Here, the ordering of the X's and Y's 
specify how the bits from dimensions $x$ and $y$ are combined to obtain a BMC $\sigma$.
The coordinates from each dimension have two bits, i.e., the \emph{bit length} $\ell$ of each dimension is 2. 
The merged bit sequence (i.e., the curve value in binary form) has $d\cdot \ell = 4$ bits. 

\begin{figure}[h]
  \centering
  \includegraphics[width=0.45\textwidth]{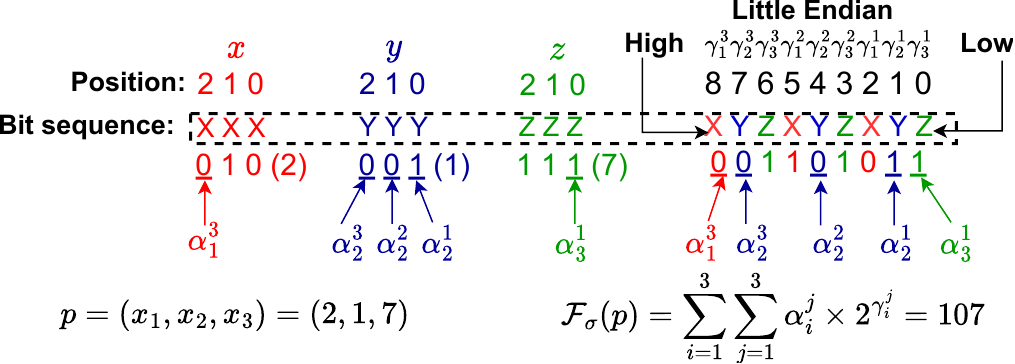}
  \caption{BMC curve value calculation ($d=3$ and  $\ell=3$).} 
  \label{fig:value_representation}
\end{figure}

The bit length $\ell$ is determined by the grid resolution, which is a system parameter. We use the same  $\ell$ for each dimension to simplify the discussion, 
and we use the little endian bit order, i.e., the rightmost bit has the lowest rank (cf. Figure~\ref{fig:value_representation}). For simplicity, we call the column indices of a point $p$ in a cell (or the cell itself) the \emph{coordinates} of $p$ (or the cell). 

\textbf{BMC value calculation.} 
Given a BMC $\sigma$, we compute the curve value of a point $p = (x_1, x_2,\dots, x_d)$ using function $\mathcal{F}_\sigma(p)$:
\begin{equation}\label{eq:fp}
\small
\mathcal{F}_\sigma(p) =\sum_{i=1}^{d}\sum_{j=1}^{\ell} \alpha_{i}^{j}\cdot 2^{\gamma_{i}^{j}}
\end{equation}
Let $x_i$ be the dimension-$i$ coordinate of $p$. In the equation, {\small$\alpha_{i}^{j} \in \{0,1\}$} is the $j$th ({\small$j\in [1, \ell]$}) bit of $x_i$, and {\small$\gamma_{i}^{j}$} is the rank of {\small$\alpha_{i}^{j}$} in the BMC.
\begin{equation}\label{eq:x_i}
\small
\sum_{j=1}^{\ell} \alpha_{i}^{j}\cdot 2^{j-1}=x_i
\end{equation}
Note that the order among the bits from the same dimension does not change when the bits are merged with those from the other dimensions to calculate $\mathcal{F}_\sigma(p)$, i.e., for bits $\alpha_i^j$ and $\alpha_i^{j+1}$, {\small$\gamma_{i}^{j} < \gamma_{i}^{j+1}$}.

For ease of discussion, we use examples with up to three dimensions $x$, $y$, and $z$.
Figure~\ref{fig:value_representation} calculates  
$\mathcal{F}_\sigma(p)$ for $p=(2,1,7)$
given $\sigma=$ XYZXYZXYZ. Here, $\alpha_{3}^{1}=1$ is the first bit value in dimension $z$, and the rank of the first (i.e., rightmost) Z bit in $\sigma$ is zero, which means $\gamma_{3}^{1}=0$. To calculate the curve value of a point for a given $\sigma$, we 
derive each $\alpha_{i}^{j}$ and $\gamma_{i}^{j}$ 
based on $x_i$ and $\sigma$, respectively. 

\vspace{1mm}
\vspace{1mm}

\textbf{BMC monotonicity.} 
The BMC value calculation process implies that any BMC is monotonic.
\begin{theorem}[Monotonicity]~\label{theorem:value_feature}
Given {\small$p_1=(x_{1, 1}, \ldots, x_{1, d})$} and {\small$p_2=(x_{2, 1}, \ldots, x_{2, d})$} then {\small$\forall i \in [1, d] ( x_{1, i} \leq x_{2, i}) \rightarrow \mathcal{F}_\sigma(p_1)\leq\mathcal{F}_\sigma(p_2)$}. 
\end{theorem}
\begin{proof}
Given {\small$x_{1, i} \leq x_{2, i}$}, 
we have {\small$\sum_{j=1}^{\ell} \alpha_{1, i}^{j}\cdot 2^{j-1} \le \sum_{j=1}^{\ell} \alpha_{2, i}^{j}\cdot 2^{j-1}$} based on Equation~\ref{eq:x_i}. The order among the bits from $x_{1, i}$ and $x_{2, i}$ do not change when they are used to calculate $\mathcal{F}_\sigma(p_1)$ and $\mathcal{F}_\sigma(p_2)$, respectively. Thus, 
{\small$\sum_{j=1}^{\ell} \alpha_{1, i}^{j}\cdot 2^{\gamma_{1, i}^{j}} \le \sum_{j=1}^{\ell} \alpha_{2, i}^{j}\cdot 2^{\gamma_{2, i}^{j}}$}.
Since this holds for any $i \in [1, d]$,
we have 
{$\small\sum_{i=1}^{d}\sum_{j=1}^{\ell} \alpha_{1, i}^{j}\cdot 2^{\gamma_{1, i}^{j}} \le \sum_{i=1}^{d}\sum_{j=1}^{\ell} \alpha_{2, i}^{j}\cdot 2^{\gamma_{2, i}^{j}}$}, i.e.,   {\small$\mathcal{F}_\sigma(p_1)\leq\mathcal{F}_\sigma(p_2)$}.
\end{proof}

\subsection{Range Querying Using a BMC} Next, we present concepts on range query processing with BMCs that will be used later to formulate query cost estimation. 
\begin{definition}[Range Query] \label{def:range_query}
Given a $d$-dimensional dataset $D$ and a 
range query  {\small$q = [x_{s, 1}, x_{e, 1}] \times [x_{s, 2}, x_{e, 2}] \times \ldots \times [x_{s, d}, x_{e, d}]$}, where {\small$[x_{s, i}, x_{e, i}]$} denotes the query range in dimension $i$, query $q$ returns all points {\small$p=(x_1, x_2,...,x_d)\in{D}$} that satisfy: {\small$\forall i\in [1,d] (x_{s, i}\leq x_i\leq x_{e, i})$}.
\end{definition}

As mentioned earlier, computing a query $q$ using different BMCs can lead to different costs. To simplify the discussion for determining the cost of a query, we use the following corollary.
\begin{corollary}\label{cor:query_section}
Given {\small$p_s=(x_{s_1}, \ldots, x_{s_d})$} and {\small$p_e=(x_{e_1}, \ldots, x_{e_d})$}, any query $q$ is bounded by the curve value range {\small$[\mathcal{F}_{\sigma}(p_s),\mathcal{F}_{\sigma}(p_e)]$}.
\end{corollary}
Corollary~\ref{cor:query_section} follows directly from the monotonicity of BMCs (Theorem~\ref{theorem:value_feature}). To simplify the discussion, we use a point $p$ and the cell that encloses $p$ interchangeably and rely on the context for disambiguation.

\textbf{Query section~\cite{QUILTS}.} A continuous curve segment in a query $q$ is called a \emph{query section}.  We denote a query section $s$ with end points $p_i$ and $p_j$    
 by {\small$[\mathcal{F}_\sigma(p_{i}), \mathcal{F}_\sigma(p_{j})]$}. Intuitively, each query section translates to a one-dimensional range query {\small$[\mathcal{F}_\sigma(p_{i}), \mathcal{F}_\sigma(p_{j})]$} on a B$^+$-tree index on dataset $D$. Thus, the number of query sections in  {\small$[\mathcal{F}_{\sigma}(p_s),\mathcal{F}_{\sigma}(p_e)]$} determines the cost of $q$. 
 
\begin{figure}[h]
	\subfloat[BMC XYXYXY~\label{fig:query_section_and_interval_1}]{
		\includegraphics[width=0.22\textwidth]{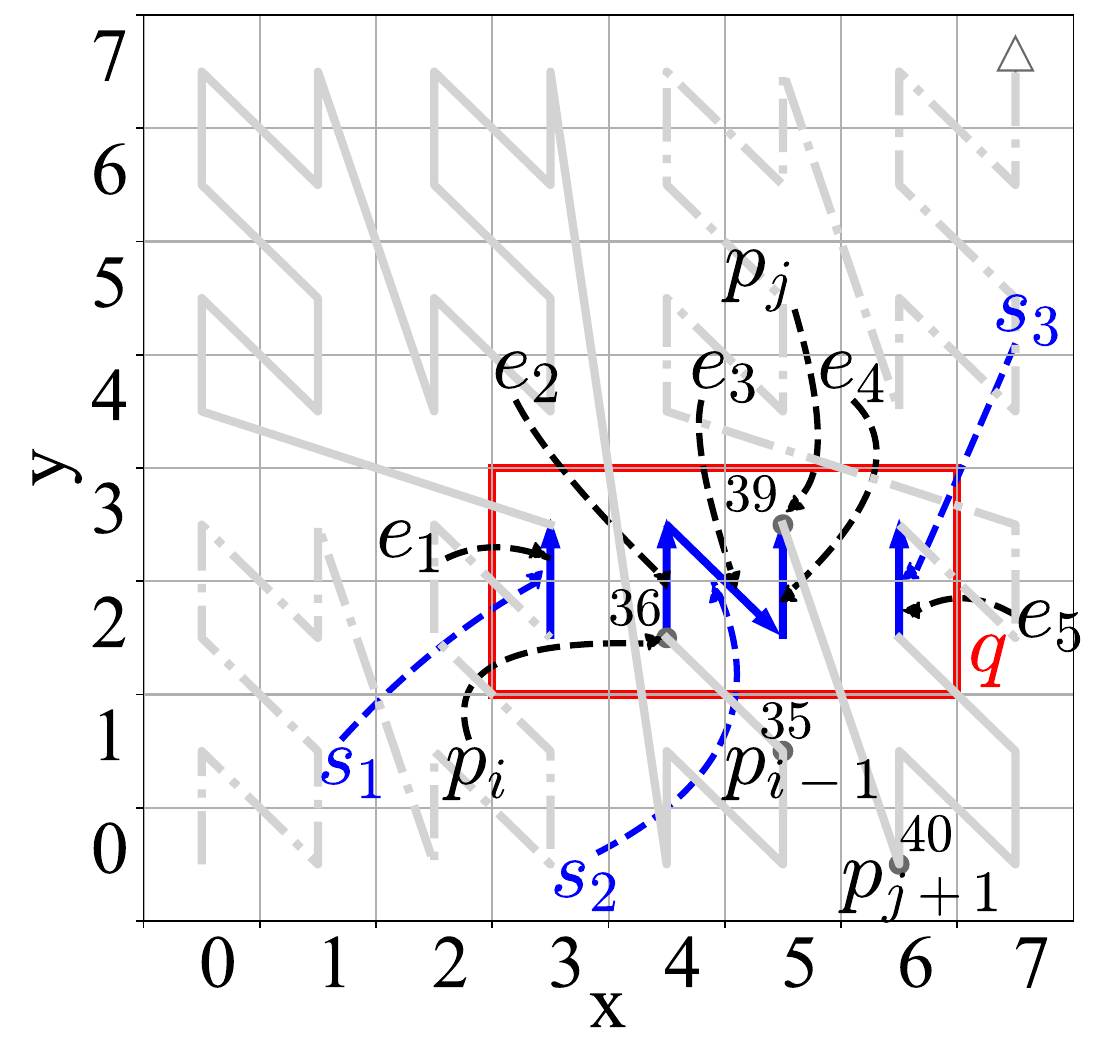}
	}
	\subfloat[BMC   YXYXYX~\label{fig:query_section_and_interval_2}]{
		\includegraphics[width=0.22\textwidth]{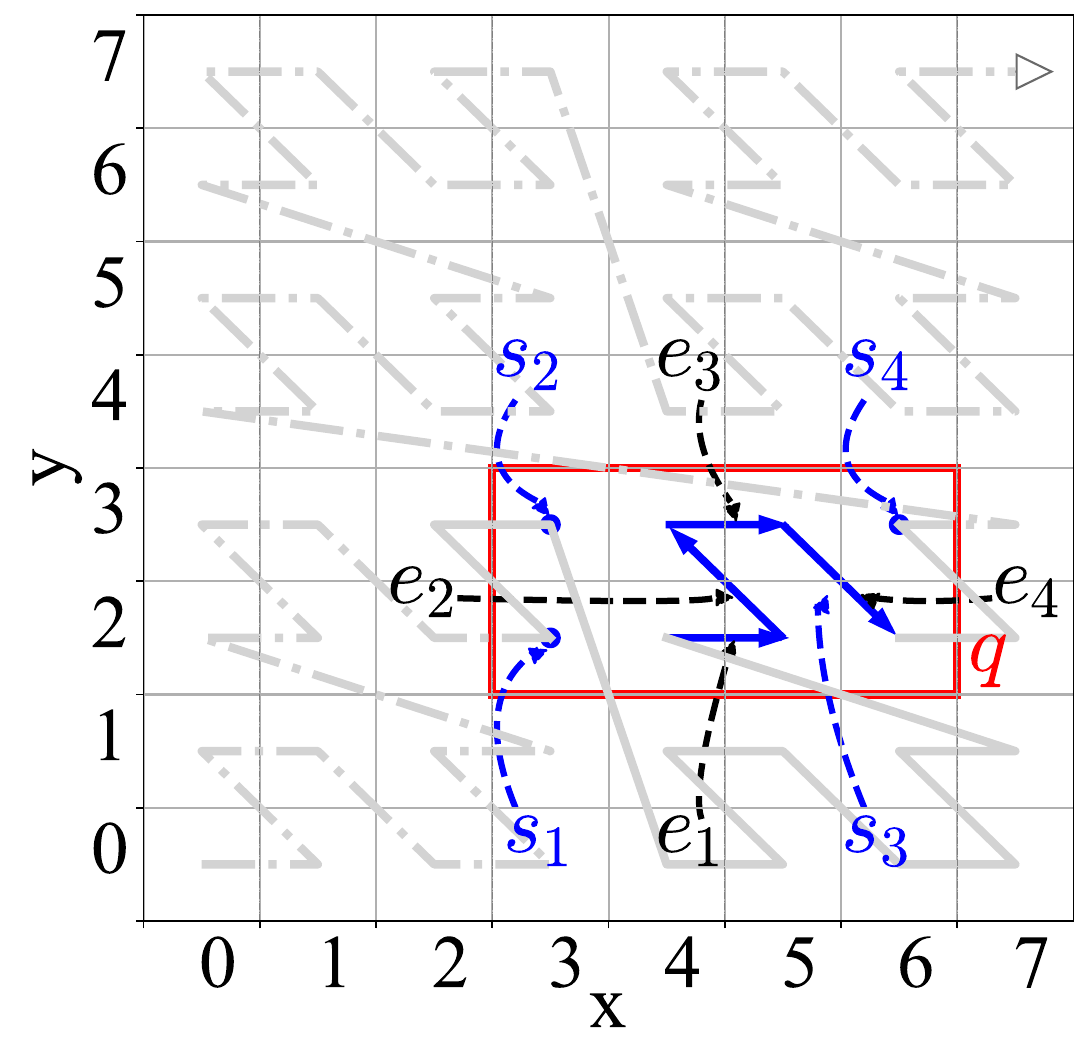}
	}
	\caption{Query sections and directed edges in BMCs.}\label{fig:query_section}
\end{figure}

\begin{example}
In Figure~\ref{fig:query_section_and_interval_1}, there are three query sections $s_1$, $s_2$, and $s_3$, with {\small$s_2 = [\mathcal{F}_\sigma(p_{i}), \mathcal{F}_\sigma(p_{j})] = [36, 39]$}. By definition, a point (cell) immediately preceding $p_i$ or succeeding $p_j$ must be outside $q$; otherwise, it is part of the query section. For example, $p_{i-1}$ ({\small$\mathcal{F}_\sigma(p_{i-1})=35$}) and $p_{j+1}$ ({\small$\mathcal{F}_\sigma(p_{j+1})=40$}) in Figure~\ref{fig:query_section_and_interval_1} are outside $q$. 
The number of query sections in $q$ varies across different BMCs, e.g., the same 
$q$ as in Figure~\ref{fig:query_section_and_interval_1} has four query sections in Figure~\ref{fig:query_section_and_interval_2}.
\end{example}
\textbf{Directed edge~\cite{SFC_theoretical_result}.} 
Query sections are composed by connecting a series of points (cells). The pair of two consecutive points $p_i$ and $p_j$ forms 
a \textit{directed edge} (denoted by $e$) 
if the curve values of $p_i$ and $p_j$ differ by one under a given $\sigma$, i.e., {\small$\mathcal{F}_\sigma(p_j) - \mathcal{F}_\sigma(p_i)=1$}. As each point is represented through a binary value, the difference occurs because
{\small$\mathcal{F}_\sigma(p_i)=\underbrace{...}_\mathit{prefix}\underline{0}\underbrace{1...1}_{K\text{ 1s}}$} and {\small$\mathcal{F}_\sigma(p_j)=\underbrace{...}_\mathit{prefix}\underline{1}\underbrace{0...0}_{K\text{ 0s}}$}, where the last $K$ ($K\geq0$) bits are changed from 1 to 0 and the ($K+1$)st bit is changed from 0 to 1.
\begin{example}
We use the binary form of two pairs of integers that form directed edges to illustrate this concept, one for $K>0$ and the other for $K=0$. First, suppose that the binary representations of {\small$\mathcal{F}_\sigma(p_i)=15$} and {\small$\mathcal{F}_\sigma(p_j)=16$} are \textbf{0}\underline{0}1111 and \textbf{0}\underline{1}0000, respectively. In this case,  four bits starting from the right (i.e., $K=4$) are changed from 1 to 0, and the fifth bit is changed from 0 to 1. The last bit \textbf{0} is the shared $\mathit{prefix}$. Second, if the binary forms of {\small$\mathcal{F}_\sigma(p_i)=16$} and {\small$\mathcal{F}_\sigma(p_j)=17$} are \textbf{01000}\underline{0} and \textbf{01000}\underline{1}, respectively, only the first bit (from the right) is changed from 0 to 1, i.e., no bits ($K=0$) are changed from 1 to 0, and the shared prefix is \textbf{01000}.
\end{example}
We explain now why the number of directed edges (denoted by  {\small$\mathcal{E}_{\sigma}(q)$}) plus the number of query sections (denoted by {\small$\mathcal{S}_{\sigma}(q)$}) in a given query $q$ yields the number of distinct points (denoted by {\small$\mathcal{V}(q)$}) in $q$. The intuition is that if $q$ consists of a single section $s$, i.e., the curve stays completely inside $s$ and {\small$\mathcal{S}_{\sigma}(q)=1$} then there are {\small$\mathcal{V}(q)-1$} directed edges connecting a given start point $p_s$ and end point $p_e$ of $s$. In other words, we obtain {\small$\mathcal{E}_{\sigma}(q) + \mathcal{S}_{\sigma}(q) = \mathcal{V}(q) - 1 + 1 = \mathcal{V}(q)$}. This is because each time a curve exits a query section $s_i$ and enters the next section $s_{i+1}$, the last point in $s_i$ becomes disconnected (minus one directed edge) but one new query section is added (plus 1 for the query section) when the curve reenters $s_{i+1}$. This leads to the following equation:
\begin{equation}~\label{eq:formula_connections}
\small
\mathcal{E}_{\sigma}(q) + \mathcal{S}_{\sigma}(q) = \mathcal{V}(q)
\end{equation}
While $\mathcal{V}(q)$ is independent of $\sigma$, the values for $\mathcal{E}_{\sigma}(q)$ and $\mathcal{S}_{\sigma}(q)$ depend on $\sigma$. For example, in Figure~\ref{fig:query_section_and_interval_1} ($\sigma=$XYXYXY), 
there are 3 query sections ($\mathcal{S}_{\sigma}(q)$) and 5 directed edges ($\mathcal{E}_{\sigma}(q)$) in $q$; in Figure~\ref{fig:query_section_and_interval_2} ($\sigma=$YXYXYX), 
there are 4 query sections and 4 directed edges in $q$. Both figures have $\mathcal{V}(q)=8$ points in $q$. Equation~\ref{eq:formula_connections} is key in computing the local cost (Section~\ref{subsec:local_cost}) of a query.

\section{Efficient BMC Cost Estimation}~\label{sec:technique}
Consider a range query $q$ with start point $p_s$ and end point $p_e$ 
and assume that dataset $D$ has been indexed with a B$^+$-tree using BMC $\sigma$. A simple query algorithm accesses the range {\small$[\mathcal{F}_{\sigma}(p_s),\mathcal{F}_{\sigma}(p_e)]$} using the B$^+$-tree, and filters any false positives not included in $q$. The query cost of $q$ then relates to the length of {\small$[\mathcal{F}_{\sigma}(p_s),\mathcal{F}_{\sigma}(p_e)]$} and the number of false positives in the range. The number of false positives in turn relates to the number of query sections in $q$. Thus, we define the cost of $q$ (when using BMC $\sigma$), denoted by $\mathcal{C}_{\sigma}(q)$, as a combination of  the length of {\small$[\mathcal{F}_{\sigma}(p_s),\mathcal{F}_{\sigma}(p_e)]$} (called the \emph{global cost}, $\mathcal{C}_{\sigma}^g(q)$) and the number of query sections (called the \emph{local cost}, $\mathcal{C}_{\sigma}^l(q)$) in $q$. 
Empirically, we find that the product of the global and the local costs best differentiates the query performance of different BMC indices, which helps identify query-optimal BMC indices (i.e., the goal of our study). Hence, we define $\mathcal{C}_{\sigma}(q)$ as: 
\begin{equation}
\small
    \mathcal{C}_{\sigma}(q)=\mathcal{C}_{\sigma}^g(q)\cdot\mathcal{C}_{\sigma}^l(q)
\end{equation} 
Note that a commonly used alternative query algorithm is to break $q$ into query sections and perform a range query on the B$^+$-tree for each such section. In this case, the local cost applies directly. The global cost, on the other hand, applies implicitly, because a larger range of {\small$[\mathcal{F}_{\sigma}(p_s),\mathcal{F}_{\sigma}(p_e)]$} implies a  higher cost to examine and uncover the query sections in the range.

Note also that the cost model of QUILTS~\cite{QUILTS} uses the product of a global and a local cost. However, its definitions of global and local costs, described in Section~\ref{sec:related_work} are different from ours.

Next, we present efficient algorithms for computing the global and local costs in 
Sections~\ref{subsec:global_cost} and~\ref{subsec:local_cost}, respectively.

\subsection{Global Cost Estimation for BMC}\label{subsec:global_cost}


As mentioned above, we define the global cost of query $q$ as the length of  {\small$[\mathcal{F}_{\sigma}(p_s),\mathcal{F}_{\sigma}(p_e)]$}.

\begin{definition}[Global Cost] \label{def:global_cost}
The global cost $\mathcal{C}_\sigma^{g}(q)$ of query $q$ under BMC $\sigma$ is the length of the curve segment from $p_s$ to $p_e$: 
\begin{equation}~\label{equ:global_cost_naive}
\small{
\mathcal{C}_\sigma^{g}(q) = \mathcal{F}_\sigma(p_e) - \mathcal{F}_\sigma(p_s) + 1=\sum_{j=1}^{d}\sum_{k=1}^{\ell} (\alpha_{e, j}^{k}-\alpha_{s, j}^{k})\cdot2^{\gamma_{j}^{k}}+1}
\end{equation}
\end{definition}
\textbf{Efficient computation.} 
Following the definition, given a set $Q$ of $n$ queries, their 
total global cost can be  calculated by visiting every query $q \in Q$ and adding up  $\mathcal{C}_\sigma^{g}(q)$. 
This naive approach takes time  proportional to the number of queries to compute. To reduce the time cost without loss of accuracy, 
we rewrite the global cost as a closed-form function for efficient computation. 
\begin{equation}\label{equ:global_cost}
\small
\begin{split}
&\mathcal{C}_\sigma^{g}(Q)=\sum_{i=1}^{n}\mathcal{C}_\sigma^{g}(q_i) = \sum_{i=1}^{n}\sum_{j=1}^{d}\sum_{k=1}^{\ell} \underbrace{({\alpha_{i, e, j}^{k}}-{\alpha}_{i, s, j}^{k})}_{\text{BMC independent}}\cdot \underbrace{2^{\gamma_{j}^{k}}}_{\text{BMC dependent}}+n\\[-2pt]
&= \sum_{j=1}^{d}\sum_{k=1}^{\ell}\underbrace{\sum_{i=1}^{n} ({\alpha}_{i, e, j}^{k}-{\alpha}_{i, s, j}^{k})}_{\text{BMC independent}}\cdot 2^{\gamma_{j}^{k}}+n=\sum_{j=1}^{d}\sum_{k=1}^{\ell}A_j^k\cdot 2^{\gamma_{j}^{k}}+n
\end{split}
\end{equation}

Here, $q_i \in Q$; {\small$\alpha_{i, s, j}^{k}$} and {\small$\alpha_{i, e, j}^{k}$}  denote the $k$th bits of the coordinates of the lower and the upper end points of $q_i$ in dimension $j$, respectively; {\small$A_j^k = \sum_{i=1}^{n} ({\alpha}_{i, e, j}^{k}-{\alpha}_{i, s, j}^{k})$}, which is BMC independent and can be calculated once by scanning the $n$ range queries in $Q$ to compute the 
gap between $p_e$ and $p_s$ on the $k$th bit of the $j$th dimension, for any BMC. Only the 
term {\small$2^{\gamma_{j}^{k}}$} is BMC dependent and must be calculated for each curve because {\small$\gamma_{i}^{j}$} represents the rank of the $j$th bit from dimension $i$ of a BMC (cf. Section~\ref{sec:bit_merging}). If the BMC $\sigma$ is changed, e.g., from XYXY\textbf{XY} to XYXY\textbf{YX}, then  {\small$\gamma_{1}^{1}=1$} and {\small$\gamma_{2}^{1}=0$} are changed to {\small$\gamma_{1}^{1}=0$} and {\small$\gamma_{2}^{1}=1$}, respectively.

\textbf{Algorithm costs.} The above property helps reduce the cost of computing the global cost when given multiple candidate BMCs. For example,
when learning the best BMC from a large volume of candidate BMCs (see Section~\ref{sec:learning_bmc}), each BMC is evaluated individually
in each iteration (Algorithm~\ref{alg:deep_Q}). 
Without an efficient cost modeling, the global cost is $O(m \cdot n \cdot d \cdot \ell)$ for $m$ candidate BMCs over $n$ queries (based on Equation~\ref{equ:global_cost_naive}).  
Based on our proposed closed form method (Equation~\ref{equ:global_cost}), after an initial $O(n)$-time scan over the $n$ queries (to compute {\small$A_j^k$}), the holistic global cost over $n$ queries can be calculated in $O(m\cdot d\cdot\ell)$ time, i.e., $O(m)$ time given constant number of dimensions $d$ and number of bits $\ell$ in each dimension.


%


\subsection{Local Cost Estimation for BMC}\label{subsec:local_cost}
The local cost measures the degree of segmentation of the curve in {\small$[\mathcal{F}_{\sigma}(p_s),\mathcal{F}_{\sigma}(p_e)]$}, which indicates the number of false positive data blocks that are retrieved unnecessarily and need to be filtered. 
We define the local cost as the number of query sections, 
following existing  studies~\cite{SFC_theoretical_result,HC_theory} that use the term ``\emph{number of clusters}'' for the same concept. 

\begin{definition}[Local Cost] \label{def:local_cost}
The local cost $\mathcal{C}_\sigma^{l}(q)$ of query $q$ under BMC $\sigma$ is 
the number query sections in $q$, i.e., $\mathcal{S}_\sigma(q)$.
\end{definition}

\textbf{Intuition.} Recall that $\mathcal{V}(q)$ is the number of distinct points in $q$. We assume one data point per cell and that every $B$ data points are stored in a block. A point is a true positive if it (and its cell) is in query $q$ and a false positive if it is outside $q$ but is retrieved by the query. If $q$ has only one query section, the largest number of block accesses is {\small$\left\lfloor (\mathcal{V}(q)-2)/B\right\rfloor + 2$}, i.e., only the first and last blocks can contain false positives (at least one true positive point in each block). In this case, the precision of the query process is at least {\small$\frac{\mathcal{V}(q)}{\mathcal{V}(q) + 2\cdot(B-1)}$}. Following the same logic, if there are $n_s$ query sections, in the worst case, each query section incurs two excess block accesses, each for a block containing only one true positive point. 
The largest number of block accesses is {\small$\left\lfloor (\mathcal{V}(q)-2 \cdot n_s)/B\right\rfloor + 2 \cdot n_s$}, and the precision is  {\small$\frac{\mathcal{V}(q)}{\mathcal{V}(q) + 2\cdot n_s\cdot(B-1)}$}.
The excess block accesses grows linearly with $n_s$. Thus, we use $n_s$ to define the local cost.

\begin{figure}[h]
  \centering
  \includegraphics[width=0.4\textwidth]{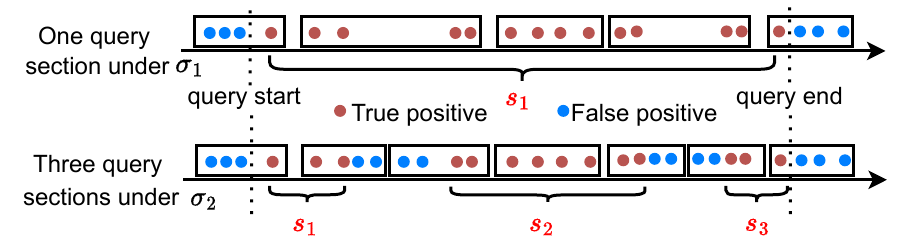}
  \caption{Query sections vs. block accesses}
  \label{fig:page_access}
\end{figure}

\begin{example}
In Figure~\ref{fig:page_access}, we order points based on BMCs $\sigma_1$ and $\sigma_2$ and place the points in blocks where $B=4$.
There are 14 true positives (i.e., {\small$\mathcal{V}(q)=14$}). There is only one query section under $\sigma_1$, which leads to a precision of $\frac{14}{5\times 4}=70\%$ for 5 block accesses, whereas $\sigma_2$ has three query sections (due to a different curve). The number of block accesses is 7, and the precision drops to $\frac{14}{7\times4}=50\%$.
\end{example}

\textbf{Efficient computation.} 
A simple way to compute the local cost of an arbitrary range query is to count the number of query sections by traversing the curve segment from $p_s$ to $p_e$, but this is also time-consuming. 
To reduce the cost,  we rewrite Equation~\ref{eq:formula_connections}  as: 
\begin{equation}~\label{equ:efficient_comp}
\mathcal{S}_\sigma(q) = \mathcal{V}(q) - \mathcal{E}_\sigma(q)
\end{equation}
Given a query $q$ and the grid resolution of the data space, it is straightforward (i.e., taking $O(d) = O(1)$ time) to compute the number of cells in $q$ (i.e., $\mathcal{V}(q)$). Then, our \textit{key insight} is that 
$\mathcal{S}_\sigma(q)$ can be computed
by counting the number of directed edges, i.e., $\mathcal{E}_\sigma(q)$, which can be done efficiently in $O(1)$ time as detailed below.  
Thus, $\mathcal{S}_\sigma(q)$ can be computed in $O(1)$ time. 


\subsubsection{Rise and Drop Patterns}\label{subsubsec:patterns}

\begin{figure}[t]
\subfloat[Rise pattern in dimension $x$ and drop pattern in dimension $y$.~\label{fig:two_patterns}]   {
    \includegraphics[width=0.23\textwidth]{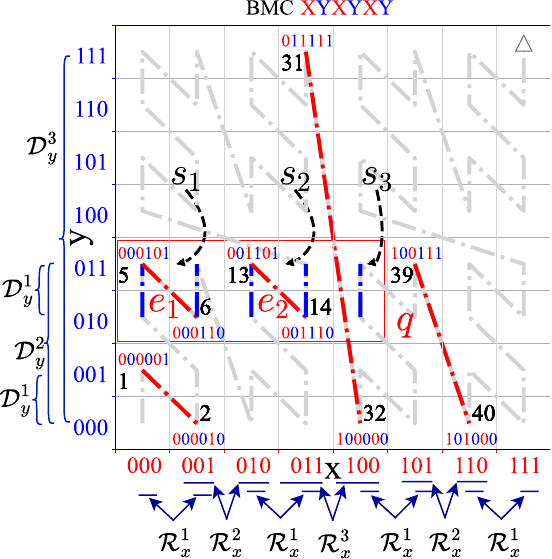}
}
\hspace{1mm}
\subfloat[Rise pattern in dimension $y$ and drop pattern in dimension $x$.~\label{fig:two_patterns_2}]   {
\includegraphics[width=0.217\textwidth]{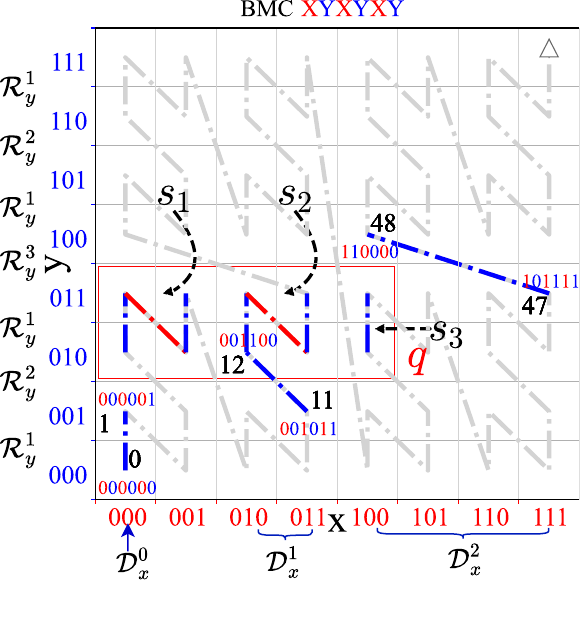}
}
\caption{Example of forming a directed edge with rise and drop patterns: for BMC XYXYXY ($d=2$ and $\ell=3$), each directed edge is formulated by a rise and a drop pattern.
}\label{fig:patterns}
\end{figure}

To compute $\mathcal{E}_\sigma(q)$ efficiently, we analyse how the bit sequence of a BMC changes from one point to another following a directed edge. 
A directed edge is formed by two consecutive points with (binary) curve values that share the same $\mathit{prefix}$, while the remaining bits are changed. We observe that different directed edges have the same shape when they share the same pattern in their changed bits, even if their prefixes are different. In Figure~\ref{fig:two_patterns}, consider edges $e_1 = (5, 6) = [\textbf{0001}\underline{01}, \textbf{0001}\underline{10}]$ and $e_2 = (13, 14) = [\textbf{0011}\underline{01}, \textbf{0011}\underline{10}]$. Both edges are in query $q$ as indicated by the red rectangle, and they share the same `\textbackslash' shape because the two rightmost bits in both cases change from ``01'' to ``10''. However, in Figure~\ref{fig:two_patterns}, edge $(1, 2) = [\textbf{0000}\underline{01}, \textbf{0000}\underline{10}]$ is not in $q$, and the prefix  (``0000'') differs from that of $e_1$ and $e_2$ above. 

A query $q$ can only contain directed edges of a few different shapes. In Figure~\ref{fig:two_patterns}, edge $(31, 32) = [\underline{011111}, \underline{100000}]$  is not in $q$, and the pattern of the changed bits differs from that of $e_1$ and $e_2$.

Note that the bits of the curve values come from the coordinates (i.e., column indices) of the two end points of a directed edge. By analyzing the bit patterns of the column indices spanned by a query $q$ in each dimension, we can count the number of directed edges that can appear in $q$. 

To generalize, recall that given a directed edge from $p_i$ to $p_j$, {\small$\mathcal{F}_\sigma(p_i)=\underbrace{...}_\mathit{prefix}\underline{0}\underbrace{1...1}_{K\text{ 1s}}$} and  {\small$\mathcal{F}_\sigma(p_j)=\underbrace{...}_\mathit{prefix}\underline{1}\underbrace{0...0}_{K\text{ 0s}}$} ($K\ge 0$) must exist where the $K$ rightmost bits are changed from 1 to 0, while the ($K+1$)st rightmost bit is changed from 0 to 1. The bits of 
{\small$\mathcal{F}_\sigma(p_i)$} and {\small$\mathcal{F}_\sigma(p_j)$} come from those of the column indices of $p_i$ and $p_j$. Thus, the $K+1$ rightmost bits changed from {\small$\mathcal{F}_\sigma(p_i)$} to {\small$\mathcal{F}_\sigma(p_j)$} must also come from  those of the column indices. In particular, there must be one dimension, where the column index has contributed $k$ ($1\le k \le K$) changed bits and one of the bits has changed from 0 to 1, while the rest dimensions contribute bits changing from 1 to 0.

Our key observation is that the bit-changing patterns across the column indices in a dimension only depend on the column indices themselves, making them \emph{BMC independent}. By pre-computing the number of bit-changing patterns that can form the $(K+1)$-bit change of a directed edge, we can derive efficiently the number of directed edges given a query $q$ and a BMC.


We summarize the bit-changing patterns to form a directed edge with two basic patterns: a \emph{rise pattern} and a \emph{drop pattern}. 


\begin{definition}[Rise Pattern]\label{def:rise_pattern}
A rise pattern {\small$\mathcal{R}_b^k$} 
of a directed edge from $p_i$ to $p_j$ 
represents a $k$-bit ($k \geq 1$) change in the dimension-$b$ coordinate of $p_i$ (i.e.,  $x_{i, b}$) to that of $p_j$ (i.e.,  $x_{j, b}$), where the rightmost $k-1$ bits are changed from 1 to 0 and the $k$th bit (from the right) is changed from 0 to 1, i.e.,  
{\small$x_{i,b}=\underbrace{...}_\mathit{prefix}\underline{0}\underbrace{1...1}_{(k - 1)\text{ 1s}}$} and  {\small$x_{j,b}=\underbrace{...}_\mathit{prefix}\underline{1}\underbrace{0...0}_{(k-1)\text{ 0s}}$}.
\end{definition}

\begin{definition}[Drop Pattern]\label{def:dropping_pattern}
A drop pattern {\small$\mathcal{D}_b^k$} 
of a directed edge from $p_i$ to $p_j$ 
represents a rightmost $k$-bit ($k \geq 0$) 1-to-0 change in the dimension-$b$ coordinate of $p_i$ (i.e.,  $x_{i, b}$) to that of $p_j$ (i.e.,  $x_{j, b}$),  
i.e., {\small$x_{i,b}=\underbrace{...}_\mathit{prefix}\underbrace{1...1}_{k\text{ 1s}}$} and {\small$x_{j,b}=\underbrace{...}_\mathit{prefix}\underbrace{0...0}_{k\text{ 0s}}$}. 
\end{definition}
Given a dimension where the coordinates use $\ell$ bits, there can be $\ell$ different rise patterns, i.e., $k \in [1, \ell]$, and there can be $\ell+1$ different drop patterns, i.e., $k\in [0, \ell]$. Note the \textit{special case} where $k=0$, i.e.,  {\small$\mathcal{D}_b^0$}, indicating no bit value drop in dimension $b$.


\begin{example}~\label{example:patterns}
In Figure~\ref{fig:two_patterns}, consider the directed edge from $p_i$ to $p_j$, where $\mathcal{F}_\sigma(p_i)=1$ (\underline{0}0\underline{0}0\underline{0}1) and $\mathcal{F}_\sigma(p_j)=2$ (\underline{0}0\underline{0}0\underline{1}0), i.e., the `
\textbackslash' segment at the bottom left. The $x$-coordinate of $p_i$changes from \underline{0}\underline{0}\underline{0} to \underline{0}\underline{0}\underline{1} to that of $p_j$ (i.e., rise pattern {\small$\mathcal{R}_x^1$}). The $y$-coordinate of $p_i$ changes from 001 to 000 to that of $p_j$ (i.e., drop pattern {\small$\mathcal{D}_y^1$}). Thus, this directed edge can be represented by a  combination of {\small$\mathcal{R}_x^1$} and {\small$\mathcal{D}_y^1$}, denoted as {\small$\mathcal{R}_x^1\oplus\mathcal{D}_y^1$}. This same combination also applies in other directed edges, such as that from {\small$\mathcal{F}_\sigma(p_i)=13$ to $\mathcal{F}_\sigma(p_j)=14$}, which is another `\textbackslash'-shaped segment. Other directed edges may use a different combination, e.g.,  {\small$\mathcal{R}_x^3\oplus\mathcal{D}_y^3$} for the one from {\small$\mathcal{F}_\sigma(p_i)=31$ to $\mathcal{F}_\sigma(p_j)=32$},  and 
{\small$\mathcal{R}_x^2\oplus\mathcal{D}_y^2$} for the one from  {\small$\mathcal{F}_\sigma(p_i)=39$ to $\mathcal{F}_\sigma(p_j)=40$}.

Figure~\ref{fig:two_patterns} has shown the rise patterns {\small$\mathcal{R}_x^k$} in dimension-$x$ and the drop patterns {\small$\mathcal{D}_y^k$} in dimension-$y$. Combining a rise and a drop pattern from these patterns forms a  directed edge in red in the figure. 

Similarly, we show in Figure~\ref{fig:two_patterns_2} the rise patterns {\small$\mathcal{R}_y^k$} in dimension-$y$ and the drop patterns {\small$\mathcal{D}_x^k$} in dimension-$x$.  Combining a rise and a drop pattern from these patterns forms a black directed edge. 
\end{example}

The \emph{pattern combination operator} `$\oplus$' applied on two (rise or drop) patterns means that the $(K+1)$-bit change of a directed edge is formed by the two patterns. 

Note also that while the rise and the drop patterns on a dimension are BMC independent, which ones that can be combined to form a directed edge is BMC dependent because different BMCs order the bits from different dimensions differently. For example, consider $\sigma=\text{X}^3\text{Y}^3\text{X}^2\text{Y}^2\text{X}^1\text{Y}^1$ (i.e., XYXYXY). From the right to the left of $\sigma$, the first rise pattern is $\mathcal{R}_x^1$. It can only be combined with drop pattern  $\mathcal{D}_y^1$, as there is just one bit $Y^1$ from dimension-$y$ to the right of $X^1$. Similarly, $\mathcal{R}_x^2$ and $\mathcal{R}_x^3$ can each be combined with  $\mathcal{D}_y^2$ and $\mathcal{D}_y^3$, respectively, i.e., all 1-bits to the right of $X^2$ and $X^3$ must be changed to 0, according to the bit-changing pattern of a directed edge. In general, for each dimension, there are only $\ell$ valid combinations of a rise and a drop pattern, and this number generalizes to $d\cdot\ell$ in a $d$-dimensional space given a BMC. 


Next, $\mathcal{E}_\sigma(q)$ can be calculated by counting the number of valid rise and drop patterns in $q$. For example, when $d=2$:
\begin{equation}\label{equ:connection}
\small
\mathcal{E}_\sigma(q) = \sum_{i=1}^{\ell}\Bigl(\mathcal{N}(\mathcal{R}_x^i)\cdot\mathcal{N}(\mathcal{D}_y^{r_y}) + \mathcal{N}(\mathcal{R}_y^i)\cdot\mathcal{N}(\mathcal{D}_x^{r_x})\Bigr)
\end{equation}
Here, $\mathcal{N}(\cdot)$ counts the number of times that a pattern occurs in $q$, and $r_x$ ($r_y$) is a parameter depending on the drop patterns that can be combined 
with $\mathcal{R}_x^i$ ($\mathcal{R}_y^i$).  In Figure~\ref{fig:patterns}, for $q = ([0,4]\times[2,3])$, there are two $\mathcal{R}_x^{1}$, one $\mathcal{R}_x^{2}$, and one $\mathcal{R}_x^{3}$, i.e., $\mathcal{N}(\mathcal{R}_x^{1})=2$, $\mathcal{N}(\mathcal{R}_x^{2})=1$, and $\mathcal{N}(\mathcal{R}_x^{3})=1$. Next, there is one  $\mathcal{D}_y^{1}$, zero $\mathcal{D}_y^{2}$, and zero $\mathcal{D}_y^{3}$ that are valid to match with these rise patterns, i.e., $\mathcal{N}(\mathcal{D}_y^{1})=1$, $\mathcal{N}(\mathcal{D}_y^{2})=0$, and $\mathcal{N}(\mathcal{D}_y^{3})=0$. Similarly, $\mathcal{N}(\mathcal{R}_y^{1})=1$, and $\mathcal{R}_y^{1}$ can be matched with $\mathcal{D}_x^{0}$ , where $\mathcal{N}(\mathcal{D}_x^{0})=5$. Recall that $\mathcal{D}_x^{0}$ is the special case with no bit value drop. It is counted as the length of the query range in dimension $x$. Overall, 
${\small \mathcal{E}_\sigma(q) = 2\times 1 +1\times 5}$. Thus, there are $10 - 7 = 3$ query sections in $q$ according to Equation~\ref{equ:efficient_comp}, which is consistent with the figure.



\textbf{Efficient counting of rise and drop patterns.} 
A rise pattern {\small$\mathcal{R}_b^{k}$} represents a change in the dimension-$b$ coordinate from {\small$x_{i, b}=a\cdot2^k+ (2^{k-1}-1)$} to {\small$x_{j, b}=a\cdot2^k + 2^{k-1}$} {\small$(a\geq0 \wedge a \in \mathbb{N})$}. 
Here, $a\cdot2^k$ is the prefix, while $2^{k-1}-1$ (i.e., {\small$\underline{0}\underbrace{1...1}_{(k - 1)\text{ 1s}}$}) and $2^{k-1}$ (i.e., {\small$\underline{1}\underbrace{0...0}_{(k - 1)\text{ 0s}}$}) represent the changed bits.  Then,   
given the data domain $[x_{s, b}, x_{e, b}]$ of dimension $b$, each pattern can be counted by calculating 
$\lfloor (x_{e, b}- 2^{k-1})/2^k\rfloor - \lceil (x_{s, b}- (2^{k-1}-1))/2^k\rceil+1$, i.e., a bound on the different values of $a$, which takes $O(1)$ time.


Similarly, a drop pattern {\small$\mathcal{D}_b^{k}$} represents a change from {\small$x_{i, b}=a \cdot 2^k + 2^k - 1$} to {\small$x_{j, b}=a \cdot 2^k + 0$} $(a\geq 0 \wedge a\in \mathbb{N})$.
Here, $a \cdot 2^k$ is the prefix, while $2^k - 1$ (i.e., {\small$\underbrace{1...1}_{k\text{ 1s}}$}) and $0$ (i.e., {\small$\underbrace{0...0}_{k\text{ 0s}}$}) represent the changed bits. 
We can count each pattern by calculating $\lfloor (x_{e, b} + 1)/2^{k}\rfloor - \lceil x_{s, b}/2^{k}\rceil$, again in $O(1)$ time.

\textbf{Generalizing to $d$ dimensions.}
As mentioned at the beginning of the subsection, a directed edge can be decomposed into a rise pattern in one dimension and drop patterns in the remaining $d-1$ dimensions. We call the set of all drop patterns in the $d-1$ dimensions a \emph{drop pattern collection}.

\begin{definition}[Drop Pattern Collection]\label{def:dropping_pattern_coll}
For a directed edge in $d$-dimensional space,
a drop pattern collection {\small$\mathcal{D}_{}^{k'}$} represents the bit combination over $d-1$ drop patterns:  {\small$\mathcal{D}_{}^{\sum_{i=1,i\neq b}^{d-1}k_i}=\biguplus_{i=1,i\neq b}^{d}\mathcal{D}_{i}^{k_i}$}  ($k' = \sum_{i=1,i\neq b}^{d}k_i=K-k$), where $b$ is the dimension with a rise pattern.
Here, `{\small$\biguplus$}' is a pattern combination operator (like  $\oplus$ above).
We note that {\small$\mathcal{D}_{}^{k'}$} and {\small$\mathcal{D}_{b}^{k}$} are interchangeable if $d=2$.
For simplicity, we call {\small$\mathcal{D}_{}^{k'}$} a drop pattern when the context eliminates any ambiguity.
\end{definition}

Now, in a $d$-dimensional data space, a directed edge can be formed by combining one rise pattern and $d-1$ drop patterns, i.e., {\small$\mathcal{R}_b^{k} \oplus \mathcal{D}_{}^{\sum_{i=1,i\neq b}^{d}k_i}=\mathcal{R}_b^{k} \oplus (\biguplus_{i=1,i\neq b}^{d}\mathcal{D}_{i}^{k_i})$} where $k'=\sum_{i=1,i\neq b}^{d}k_i$.
Equation~\ref{equ:connection} is then rewritten as: 
\begin{equation}\label{equ:connection_md}
\small
\mathcal{E}_\sigma(q) = \sum_{j=1}^{d}\sum_{i=1}^{\ell}\mathcal{N}(\mathcal{R}_j^i)\cdot\mathcal{N}(\mathcal{D}^{r})
\end{equation}
Here, the value of parameter $r$ depends on the number of drop patterns that can be combined with $\mathcal{R}_j^i$.

\subsubsection{Pattern Tables}\label{subsubsec:connection_tables}
We have shown how to compute the local cost of a query efficiently. 
Given a set $Q$ of $n$ range queries ($q_i \in Q$), their total local cost based on Definition~\ref{def:local_cost} is:
\begin{equation}
\label{equ:local_cost}
\small
\mathcal{C}_\sigma^{l}(Q)=\sum_{i=1}^{n}\mathcal{C}_\sigma^{l}(q_i)=\sum_{i=1}^{n}\mathcal{V}(q_i) - \sum_{i=1}^{n}\mathcal{E}_\sigma(q_i)
\end{equation}
This cost takes $O(n)$ time to compute. Given $m$ BMCs, computing their respective total local costs $\mathcal{C}_\sigma^{l}(Q)$ takes $O(m\cdot n)$ time.
As {\small$\sum_{i=1}^{n}\mathcal{V}(q_i)$} is independent of the BMCs, it can be computed once by performing an $O(n)$-time scan over $Q$. The computational bottleneck for $m$ BMCs is then the computation of {\small$\sum_{i=1}^{n}\mathcal{E}_\sigma(q_i)$}.

We eliminate this bottleneck by introducing a look-up table called a \textit{pattern table} that stores pre-computed numbers of rise-and-drop pattern combinations to form the directed edges at different locations, which are BMC-independent. Since each directed edge is a combination of a rise pattern in some dimension $b$ and $d-1$ drop patterns, we proceed to show how to pre-compute $d$ pattern tables, each recording the rise patterns of a dimension. 

\vspace{1mm}
\begin{table}[h]
    \centering
    \small
    \setlength{\tabcolsep}{2
    pt}
    \caption{Pattern table $\mathit{Table}^b$ for dimension $b$ using $\ell$ bits on each dimension.}
    \label{tab:table_rise_drop}
    \begin{tabular}{c|c|c|c|c}
      \toprule  
       & $\mathcal{D}^0$ & $\mathcal{D}^1$ &  $\cdots$ 
      & $\mathcal{D}^{\ell\cdot(d-1)}$ \\
      \midrule
      \midrule
      $\mathcal{R}_{b}^1$ & $\mathcal{N}(\mathcal{R}_b^1)\cdot \mathcal{N}(\mathcal{D}^0)$ & $\mathcal{N}(\mathcal{R}_b^1)\cdot \mathcal{N}(\mathcal{D}^1)$ & 
      $\cdots$
     & $\mathcal{N}(\mathcal{R}_b^1)\cdot \mathcal{N}(\mathcal{D}^{\ell\cdot(d-1)})$\\
      \hline
      $\mathcal{R}_{b}^2$ & $\mathcal{N}(\mathcal{R}_b^2)\cdot \mathcal{N}(\mathcal{D}^1) $ & $\mathcal{N}(\mathcal{R}_b^2)\cdot \mathcal{N}(\mathcal{D}^2) $&
      $\cdots$ & $\mathcal{N}(\mathcal{R}_b^2)\cdot \mathcal{N}(\mathcal{D}^{\ell\cdot(d-1)})$\\
      \hline
    $\cdots$ & $\cdots$ &$\cdots$& $\cdots$& $\cdots$\\
    \hline
      $\mathcal{R}_{b}^{\ell}$ & $\mathcal{N}(\mathcal{R}_b^\ell)\cdot \mathcal{N}(\mathcal{D}^0)$ & $\mathcal{N}(\mathcal{R}_b^\ell)\cdot \mathcal{N}(\mathcal{D}^1) $ &
      $\cdots$
      & $\mathcal{N}(\mathcal{R}_b^\ell)\cdot \mathcal{N}(\mathcal{D}^{\ell\cdot(d-1)})$\\
      \bottomrule
    \end{tabular}
  \end{table}

\begin{definition}[Pattern Table]\label{def:connection_table}
The pattern table for dimension $b$, denoted by {\small$\mathit{Table}^b$}, contains $\ell$ rows, each corresponding to a rise pattern in the dimension,  and $\ell\cdot(d-1)+1$ columns, each corresponding to a drop pattern in the other $d-1$ dimensions. 
As shown in Table~\ref{tab:table_rise_drop}, the value in row $i$ and column $j$ is the product of 
the numbers of rise pattern $\mathcal{R}_b^i$ and drop pattern $\mathcal{D}^j$.
\end{definition}
\vspace{-2mm}

There is a total of $\ell\cdot(d-1)+1$ drop patterns in the $d-1$ dimensions because there are $\ell\cdot(d-1)$ bits in those dimensions, i.e., $k' \in [0, \ell\cdot(d-1)]$ for   $\mathcal{D}^{k'}$. Further, since the rise and drop patterns correspond to only the bit sequences in each dimension and not the curve values, the values in the pattern tables can be computed once given a set of queries $Q$ and can then be reused across local cost estimation for different BMCs. 
Algorithm~\ref{alg:gen_table} summarizes the steps to compute pattern table {\small$\mathit{Table}^b$} based on its definition.

\begin{algorithm}[h]
\setstretch{0.7}
    \begin{small}
      \caption{Generate pattern table (GPT)} \label{alg:gen_table}
      \KwIn{Query set $Q$, target dimension $b$, data dimensionality $d$, number of bits per dimension $\ell$}
      \KwOut{Pattern table $\mathit{Table}^b$}
      Initialize an $\ell \times (\ell\cdot(d-1) + 1)$ table $\mathit{Table}^b$\;
      \For{$q\in Q$} {
        \For{$i\in [1, \ell]$} {
            \For{$j\in [0, \ell\cdot(d-1)]$} {
                $\mathcal{N}(\mathcal{R}_b^i)\leftarrow$ count the number of $\mathcal{R}_b^i$ in $q$\; 
                $\mathcal{N}(\mathcal{D}^j)\leftarrow$ count the number of $\mathcal{D}^j$ in $q$\; 
                $\mathit{Table}^b[i][j] \leftarrow \mathit{Table}^b[i][j]+ \mathcal{N}(\mathcal{R}_b^i) \cdot \mathcal{N}(\mathcal{D}^j)$\;
            }
        }
      }
      \Return $\mathit{Table}^b$\;
    \end{small}
\end{algorithm}
\vspace{-2mm}

\begin{figure}[h]
	\subfloat[Six directed edges ($\sigma=$ XYXYXY)~\label{fig:value_patterns_0}]{
		\includegraphics[width=0.225\textwidth]{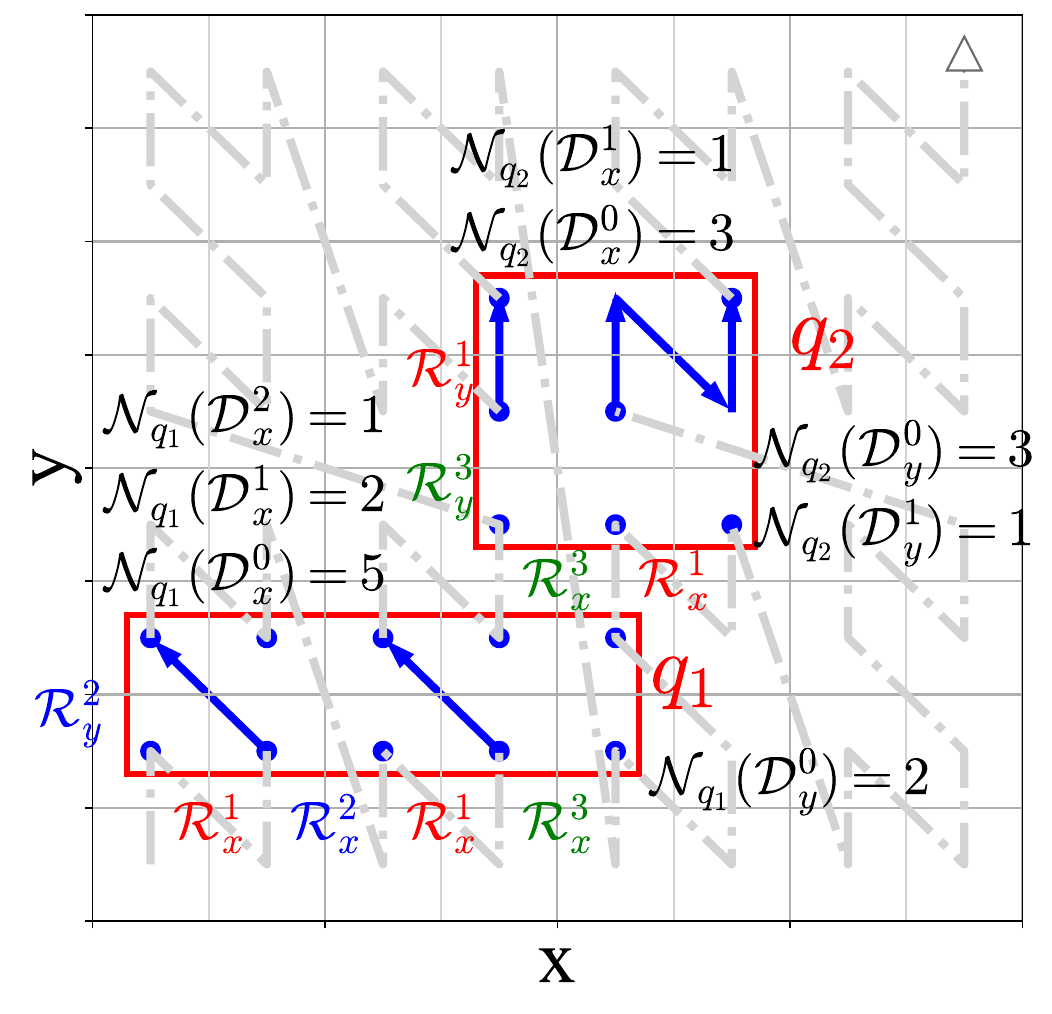}
	}
	\subfloat[Nine directed edges ($\sigma=$ YXYXYX)~\label{fig:value_patterns_1}]{
		\includegraphics[width=0.22\textwidth]{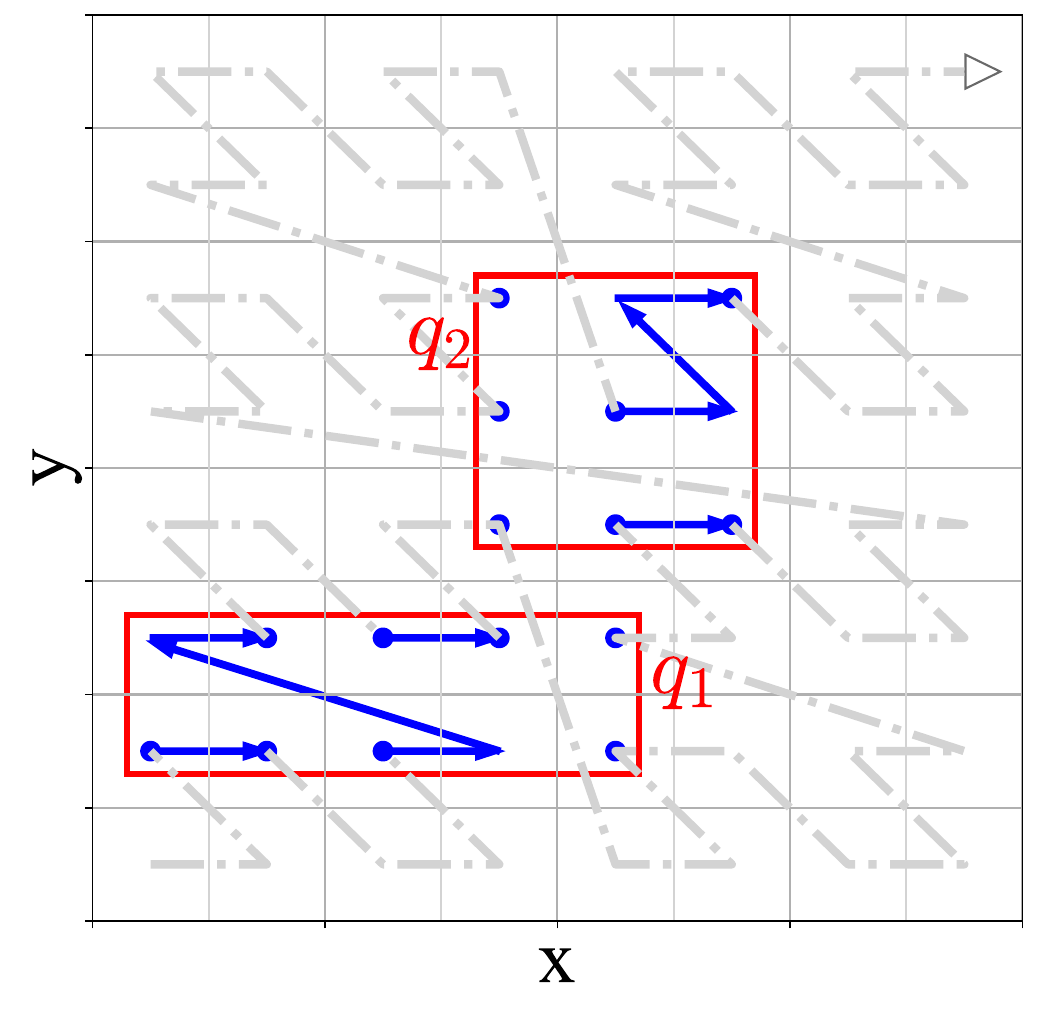}
	}
	\caption{Rise and drop pattern counting example ($d=2, \ell=3$). The results are shown in 
	pattern tables in Tables~\ref{tab:table_1_demo} and~\ref{tab:table_2_demo}. }\label{fig:connection_table_example} 
\end{figure}

\begin{example}~\label{example:gen_table}
In Figure~\ref{fig:value_patterns_0}, we show two queries $q_1$ and $q_2$, and the pattern tables $\mathit{Table}^x$ and $\mathit{Table}^y$ are shown in Tables~\ref{tab:table_1_demo} and~\ref{tab:table_2_demo}, respectively. In the tables, we use `$+$' to denote summing up the pattern table cell values (i.e., {\small$\mathcal{N}(\mathcal{R}_b^i) \cdot \mathcal{N}(\mathcal{D}^j)$}, and {\small$\mathcal{N}(\mathcal{D}^j)$} is {\small$\mathcal{N}(\mathcal{D}_x^j)$} or {\small$\mathcal{N}(\mathcal{D}_y^j)$}) computed for $q_1$ and $q_2$. For example, in $q_1$, {\small$\mathcal{N}(\mathcal{R}_x^1)=2$} (the two {\small$\mathcal{R}_x^1$} are labeled for $q_1$ in Figure~\ref{fig:value_patterns_0}) and {\small$\mathcal{N}(\mathcal{D}_y^0)=2$} (the value range of $q_1$ in dimension $y$ is 2).
Meanwhile, in $q_2$, {\small$\mathcal{N}(\mathcal{R}_x^1)=1$} (one {\small$\mathcal{R}_x^1$} is labeled for $q_2$ in Figure~\ref{fig:value_patterns_0})  
and {\small$\mathcal{N}(\mathcal{D}_y^0)=3$} (the value range of $q_2$ in dimension $y$ is 3).
Thus, in $\mathit{Table}^x$,
the cell $\mathit{Table}^x[1][0]$ (corresponding to {\small$\mathcal{R}_{x}^1 \oplus \mathcal{D}_y^0$}) is the sum of  {\small$\mathcal{N}(\mathcal{R}_x^1)\cdot \mathcal{N}(\mathcal{D}_y^0)$} in $q_1$ and $q_2$, i.e.,  $4 + 3$.
\end{example}

\vspace{-5mm}

\vspace{1mm}
\begin{table}[h]
\centering
   \begin{minipage}[h]{0.22\textwidth}
\centering
\small
    \captionof{table}{$\mathit{Table}^x$}
    \label{tab:table_1_demo}
    \setlength{\tabcolsep}{4pt}
    \begin{tabular}{c|c|c|c|c}
      \toprule  
      & $\mathcal{D}_y^0$ & $\mathcal{D}_y^1$ & $\mathcal{D}_y^2$&$\mathcal{D}_y^{3}$ \\
      \midrule
      \midrule
      $\mathcal{R}_{x}^1$ & \underline{4 + 3} & \uwave{0 + 1} & 0 + 0 & 0 + 0\\
      $\mathcal{R}_{x}^2$ & {2 + 0}  & \underline{0 + 0} & \uwave{0 + 0} & 0 + 0\\
      $\mathcal{R}_{x}^{3}$ & {2 + 3}  & {0 + 1} & \underline{0 + 0} & \uwave{0 + 0}\\
      \bottomrule
    \end{tabular}
\end{minipage}
    \hspace{2mm}
 \begin{minipage}[h]{0.22\textwidth}
\centering
    \small
    \captionof{table}{$\mathit{Table}^y$}
    \label{tab:table_2_demo}
    \setlength{\tabcolsep}{4pt}
    \begin{tabular}{c|c|c|c|c}
      \toprule  
      & $\mathcal{D}_x^0$ & $\mathcal{D}_x^1$ & $\mathcal{D}_x^2$&$\mathcal{D}_x^{3}$ \\
      \midrule
      \midrule
      $\mathcal{R}_{y}^1$ & \uwave{0 + 3} & \underline{0 + 1} & 0 + 0 & 0 + 0\\
      $\mathcal{R}_{y}^2$ & {5 + 0}  & \uwave{2 + 0} & \underline{1 + 0} & 0 + 0\\
      $\mathcal{R}_{y}^{3}$ & 0 + 3 & {0 + 1} & \uwave{0 + 0} & \underline{0 + 0}\\
      \bottomrule
    \end{tabular}
\end{minipage}
\end{table}

\subsubsection{Local Cost Estimation with Pattern Tables}\label{subsubsec:read_loop_up_tables}

Next, we describe how to derive the number of directed edges (and hence compute the total local cost) given the $d$ pattern tables for $n$ queries. 

Algorithm~\ref{alg:read_table} shows how to compute the local cost using the pattern tables. 
Each dimension $j$ is considered for the rise patterns (Line 2). Then, we consider each rise pattern in the dimension, i.e., each row $i$ in $\mathit{Table}^j$ (Line 3). We locate 
the corresponding drop pattern (i.e., the table column index) based on $i$ and a given BMC $\sigma$, which is done by the 
\texttt{get\_col} function (Line 4). Then, we add the cell value to the number of directed edges $\mathcal{E}_\sigma$ (Line 5).
Note that all $\ell$ rise patterns in each dimension are considered because a BMC has $\ell$ bits on each dimension, which can all be the bit that changes from 0 to 1. 
We return the total local cost by subtracting the total number of directed edges from the total number of cells in $Q$.

\begin{algorithm}[h]
\setstretch{0.6}
    \begin{small}
      \caption{Compute local cost with pattern tables} \label{alg:read_table}
      \KwIn{BMC $\sigma$, data dimensionality $d$, number of bits per dimension $\ell$, all pattern tables $\mathit{Table}^j$, total number of cells in the queries $\mathcal{V}$}
      \KwOut{Total local cost of $n$ queries}
      $\mathcal{E}_\sigma \leftarrow 0$\;
      \For{$j\in [1, d]$} {
          \For{$i\in [1, \ell]$} {
            $col\leftarrow\texttt{get\_col}(\sigma, i, j)$\;
            $\mathcal{E}_\sigma \leftarrow \mathcal{E}_\sigma + \mathit{Table}^j[i][col]$\;
          }
      }
      \Return $\mathcal{V}-\mathcal{E}_\sigma$\;
    \end{small}
\end{algorithm}

\begin{example}
Based on Example~\ref{example:gen_table}, 
given BMC XYXYXY, from {\small$\mathit{Table}^x$}, we read cells ({\small$\mathcal{R}_1^1$, $\mathcal{D}_2^1$}), ({\small$\mathcal{R}_1^2$, $\mathcal{D}_2^2$}), and ({\small$\mathcal{R}_1^3$, $\mathcal{D}_2^3$}), i.e., the cells with \say{wavy} lines. Similarly, we read the cells with \say{wavy} lines from {\small$\mathit{Table}^y$}. These cells sum up to 6, which is the number of directed edges (segments with arrows) in Figure~\ref{fig:value_patterns_0}. 
Similarly, the cells relevant to BMC YXYXYX are underlined, which yields a total of
 nine directed edges in Figure~\ref{fig:value_patterns_1}.
\end{example}

\textbf{Algorithm costs.}
In general, for each rise pattern, the total number of  possible drop pattern combinations is {\small$(\ell+1)^{d-1}$} based on Definition~\ref{def:dropping_pattern_coll}. 
The time complexity of generating the $d$ pattern tables is {\small$O(d\cdot\ell\cdot(\ell+1)^{d-1})$}, where $d$ denotes the number of dimensions, $\ell$ denotes the number of rows, and {\small$(\ell+1)^{d-1}$} denotes the accumulated number of drop patterns (equal to $(\ell+1)$ when $d=2$).
After initialization, the retrieval time complexity of pattern tables is {\small$O(d\cdot\ell) = O(1)$}, i.e.,  we retrieve $\ell$ cells from each table.

We generate $d$ pattern tables, each with {\small$\ell\cdot(\ell+1)^{d-1}$} keys. Thus, the space complexity for the pattern tables is  {\small$O(d\cdot\ell\cdot(\ell+1)^{d-1})$}.  
For example, when $d=3$ and $\ell=32$, all the tables take 1.6 MB (1.2 MB for keys and 0.4 MB for values). 




\section{Cost Estimation-Based BMC Learning}~\label{sec:learning_bmc} 
Next, powered by our efficient cost estimations, we aim to find the optimal BMC $\sigma_{opt}$ that minimizes the costs of a set of queries $Q$ on a dataset $D$. 
While using BMCs reduces the number of curve candidates from {\small$({2^\ell})^d!$} to  {\small$\frac{(d\cdot\ell)!}{(\ell!)^d}$} 
(Section~\ref{sec:intro}), it is still non-trivial to find the optimal BMC from the {\small$\frac{(d\cdot\ell)!}{(\ell!)^d}$} candidates. 
We present an efficient learning-based algorithm named \emph{\method} for this search.

\textbf{Problem transformation}. 
%
Starting from any random BMC $\sigma$, the process to search for  $\sigma_{opt}$ can be seen as a bit-swapping process, until every bit falls into its optimal position, assuming an oracle to guide the bit-swapping process. 

To reduce the search space, we impose two constraints on the bit swaps: (a)  we only swap two adjacent bits each time, and (b) two bits from the same dimension cannot be swapped (which guarantees valid BMCs after swaps,  cf.~Section~\ref{sec:bit_merging}). Any bit then takes at most {\small$(d-1)\cdot \ell$} swaps to reach its optimal position, when such a position is known. Given 
$d\cdot \ell$ bits, at most {\small$d\cdot(d-1)\cdot\ell^2$} swaps are needed to achieve the optimal BMC guided by an oracle.

In practice, an ideal oracle is unavailable. Now the problem becomes how to run the bit swaps without an ideal oracle. There are two approaches: (a) run a random swap (i.e., \emph{exploration}) each time and keep the result if it reduces the query cost, and (b) select a position that leads to the largest query cost reduction each time (i.e., \emph{exploitation}). Using either approach yields local optima. We integrate both approaches by leveraging \emph{deep reinforcement learning} (DRL) to approach a global optimum, since DRL aims to maximize a long-term objective~\cite{ADI} and balance exploration and exploitation.

\textbf{BMC learning formulation.} 
We formulate BMC learning as a DRL problem:
    (1) State space $\mathcal{S}$, where a state (i.e., a BMC) $\sigma_t \in \mathcal{S}$ at time step $t$ is a vector $\langle \sigma_t[d\cdot \ell],\sigma_t[d\cdot \ell-1],\ldots,\sigma_t[1]\rangle$, and $\sigma_t[i]$ is the $i$th bit.  
   For example, if $\sigma_t=$XYZ, $\sigma_t[3]$=X, $\sigma_t[2]$=Y, and $\sigma_t[1]$=Z.
   (2) Encoding function $\phi(\cdot)$, which encodes a BMC to fit the model input. We use one-hot encoding. For example, X, Y, and Z can be encoded into $[0,0,1]$, $[0,1,0]$, and $[1,0,0]$, respectively, and XYZ  by $[0,0,1,0,1,0,1,0,0]$.
    (3) Action space $\mathcal{A}$, where an action $a \in \mathcal{A}$ is the position of a bit to swap. When the $a$th bit is chosen, we swap it with the $(a+1)$st bit (if $a+1 \le d\cdot\ell$).
    Thus, $\mathcal{A}=\{a\in \mathbb{Z}:1\leq a\leq d\cdot\ell - 1\}$.
    (4) Reward $r$: $\mathcal{S} \times \mathcal{A} \times \mathcal{S} \to r$, which is the query cost reduction when reaching a new BMC $\sigma_{t+1}$ from $\sigma_{t}$. 
    Since an oracle is unavailable, we use our cost model to estimate the query cost of a BMC. 
    The reward $r_t$ at step $t$ is calculated as $r_t =(\mathcal{C}_{\sigma_{t}} - \mathcal{C}_{\sigma_{t+1}}) / \mathcal{C}_{\sigma_{1}}$, where $\mathcal{C}_{\sigma_{t}}=\mathcal{C}_{\sigma_{t}}^g(Q)\cdot\mathcal{C}_{\sigma_{t}}^l(Q)$ is the cost of $\sigma_{t}$ estimated by Equation~\ref{equ:global_cost} and Algorithm~\ref{alg:read_table}. 
    (5) Parameter $\epsilon$, which balances 
    exploration and exploitation to avoid local optima. 


Based on this formulation, we use \emph{deep Q-learning}~\cite{DQN} in our \method\ algorithm to learn a query-efficient BMC index. 

\textbf{The \method\  algorithm.} 
We summarize \method\ in Algorithm~\ref{alg:deep_Q} where the input 
$\sigma_1$ can be any initial BMC, e.g., a ZC. 
The key idea of \method\ is to learn a policy $\pi: \mathcal{S} \rightarrow \mathcal{A}$ that guides the position selection for a bit swap given a status, to maximize a value function $\mathtt{Q^{*}}(\phi(\sigma_t), a)$ (i.e., the reward) at each step $t$.
Such a policy $\pi$ can be learned by training a model (a \emph{deep Q-network}, DQN) with parameters $\theta$ over existing  ``\emph{experience}'' (previously observed state transitions and their rewards), which is used to predict the position $a$ to maximize the value function (i.e., {\small$\max_{a}\mathtt{Q^{*}}(\phi(\sigma_t), a; \theta)$}). After a number of iterations, the learned BMC $\sigma_{opt}^*$ is expected to approach $\sigma_{opt}$, which is returned as the algorithm output.

We initialize a storage $MQ$ to store the latest $N_{MQ}$ bit-swapping records (i.e., the experience, Line~1). 
We learn to approach $\sigma_{opt}$ with $M$ episodes and $T$ steps per episode (Lines 2 and 3). In each episode, we start with $\sigma_1$ encoded by $\phi(\cdot)$.  
To select a swap position $a_t$ at step $t$, we generate a random number in $[0, 1]$, if the number is greater than $\epsilon$, we randomly select a position $a_t$, otherwise, we set $a_t$ as the position with the highest probability to obtain a  maximal reward, i.e., 
{\small$\max_{a}\mathtt{Q^{*}}(\phi(\sigma_t), a; \theta)$} (Line 4).
The prediction is based on the current state $\sigma_t$ and model weights $\theta$. 
We execute $a_t$ ({\small$\mathtt{E}(\sigma_t,a_t)$} at Line 5) and compute reward $r_t$ using our cost model (Line 6). 
We record the new transition in $MQ$ and train the DQN (i.e., update $\theta$) over sampled data in $MQ$ (Lines 7 and 8).
The training uses gradient descent to minimize a loss function {\small${L_t}(\theta_t)=\mathbb{E}_{\phi(\sigma),a\sim\rho(\cdot)}\left[(y_t - Q(\phi(\sigma),a;\theta_{t}))^2\right]$}
where $y_t$ is the target from iteration $t$ and $\rho(\cdot)$ is the action distribution~\cite{DQN}.
We use $\sigma_{opt}^*$ to record the new BMC from each swap (Line~9), which is returned in the end (Line~10).

\begin{algorithm}[h]
\setstretch{0.7}
    \begin{small}
      \caption{Learn BMC (\method)} \label{alg:deep_Q}
      \KwIn{Initial BMC $\sigma_1$}
      \KwOut{A query-efficient BMC  $\sigma_{opt}^*$}
      Initialize replay memory $MQ$ with capacity $N_{MQ}$\;
      \For{$episode\in [1,M]$} {
           \For{$t\in [1,T]$} {
                With probability $\epsilon$ select a random position $a_t$, or $a_t \leftarrow \max_{a}\mathtt{Q^{*}}(\phi(\sigma_t), a; \theta)$\;
                $\sigma_{t+1}\leftarrow \mathtt{E}(\sigma_t,a_t)$\;
                Compute reward $r_t$\;
                Store transition $(\phi(\sigma_{t}),a_t,r_t,\phi(\sigma_{t+1}))$ in $MQ$\;
                Train model $\theta$ with sampled transitions from $MQ$\;
                $\sigma_{opt}^* \leftarrow \sigma_{t+1}$\;
           }
      }
      \Return $\sigma_{opt}^*$\;
    \end{small}
\end{algorithm}


\begin{figure}[h]
	\subfloat[YXXYYX, $\mathcal{C}_1=175$~\label{fig:RL_0}]{
		\includegraphics[width=0.15\textwidth]{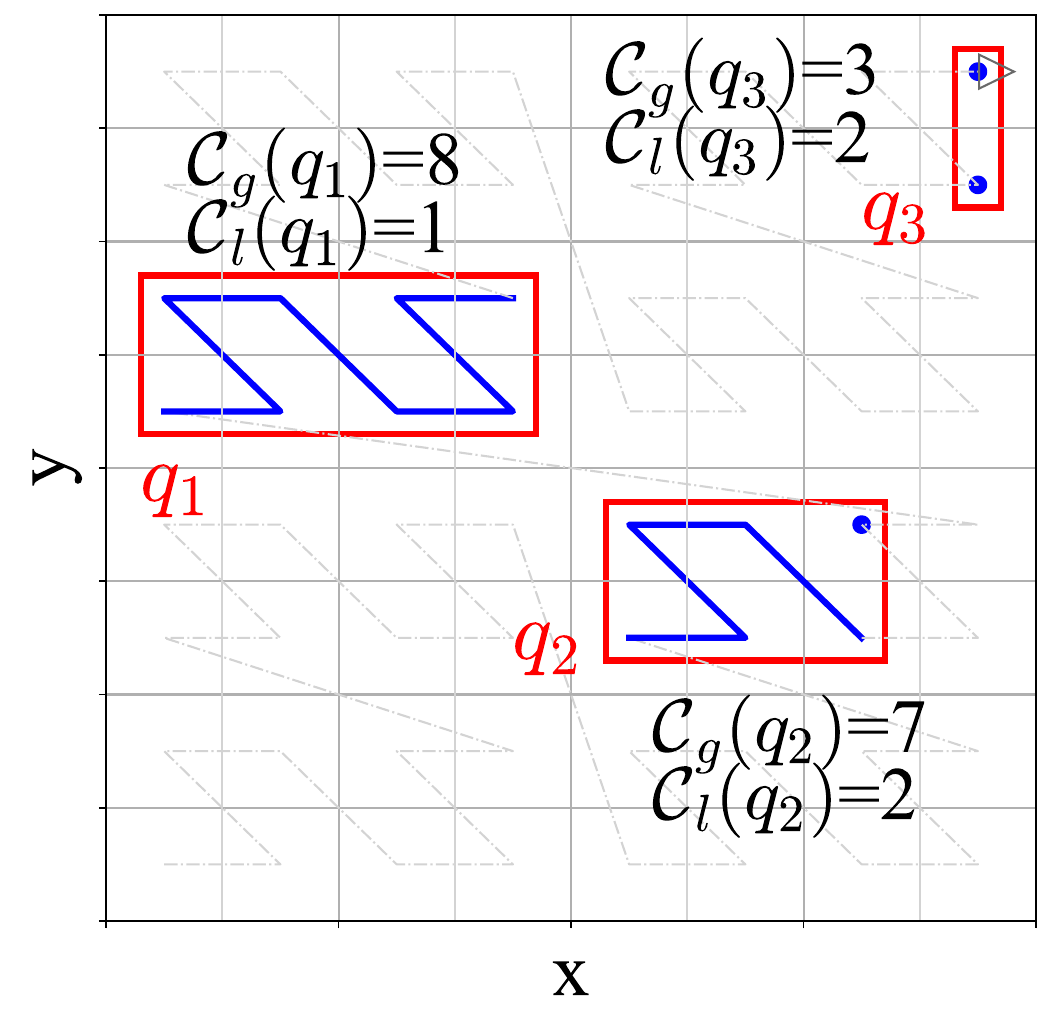}
  \hspace{-2mm}
	}
	\subfloat[YX{\color{red}YX}YX, $\mathcal{C}_2=90$~\label{fig:RL_1}]{
		\includegraphics[width=0.15\textwidth]{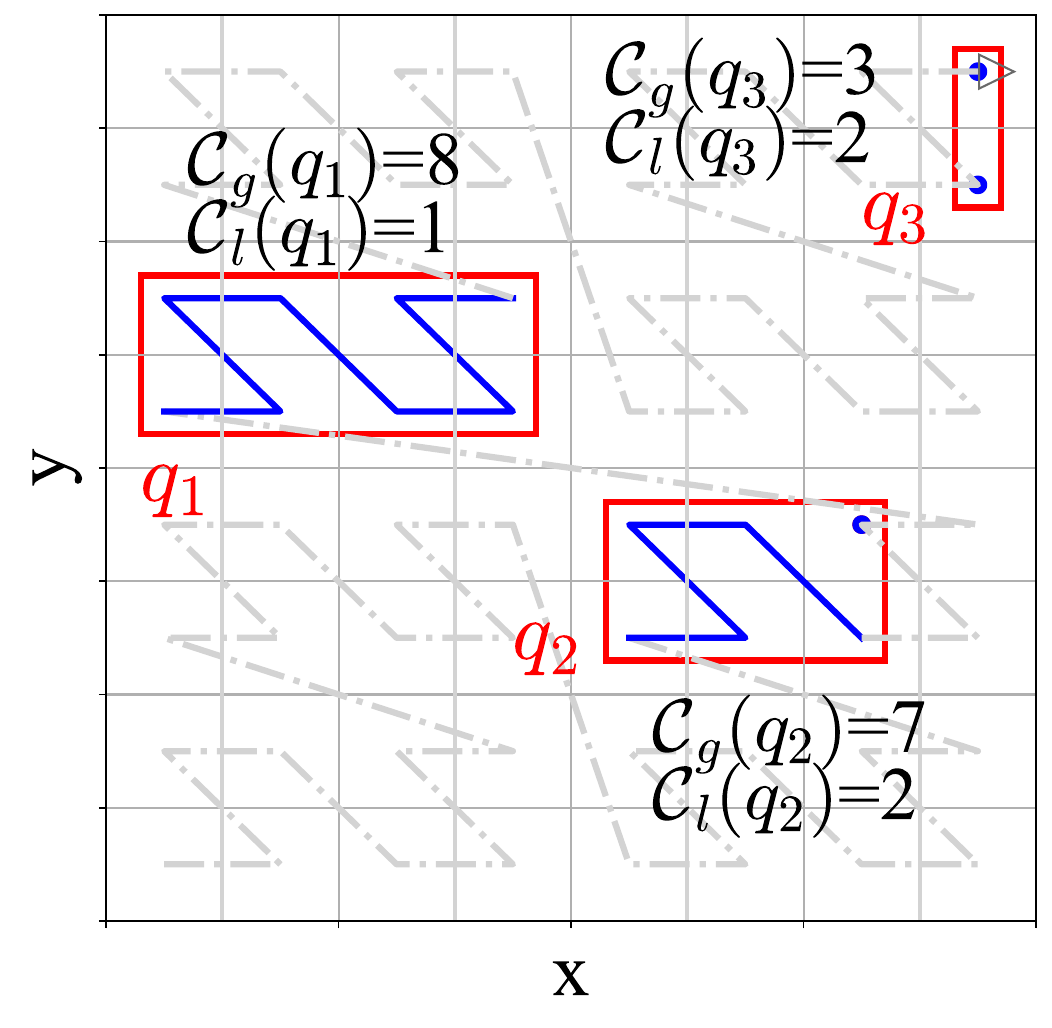}
  \hspace{-2mm}
	}
	\subfloat[YX{\color{red}YX}{\color{blue}XY}, $\mathcal{C}_3=48$~\label{fig:RL_2}]{
		\includegraphics[width=0.15\textwidth]{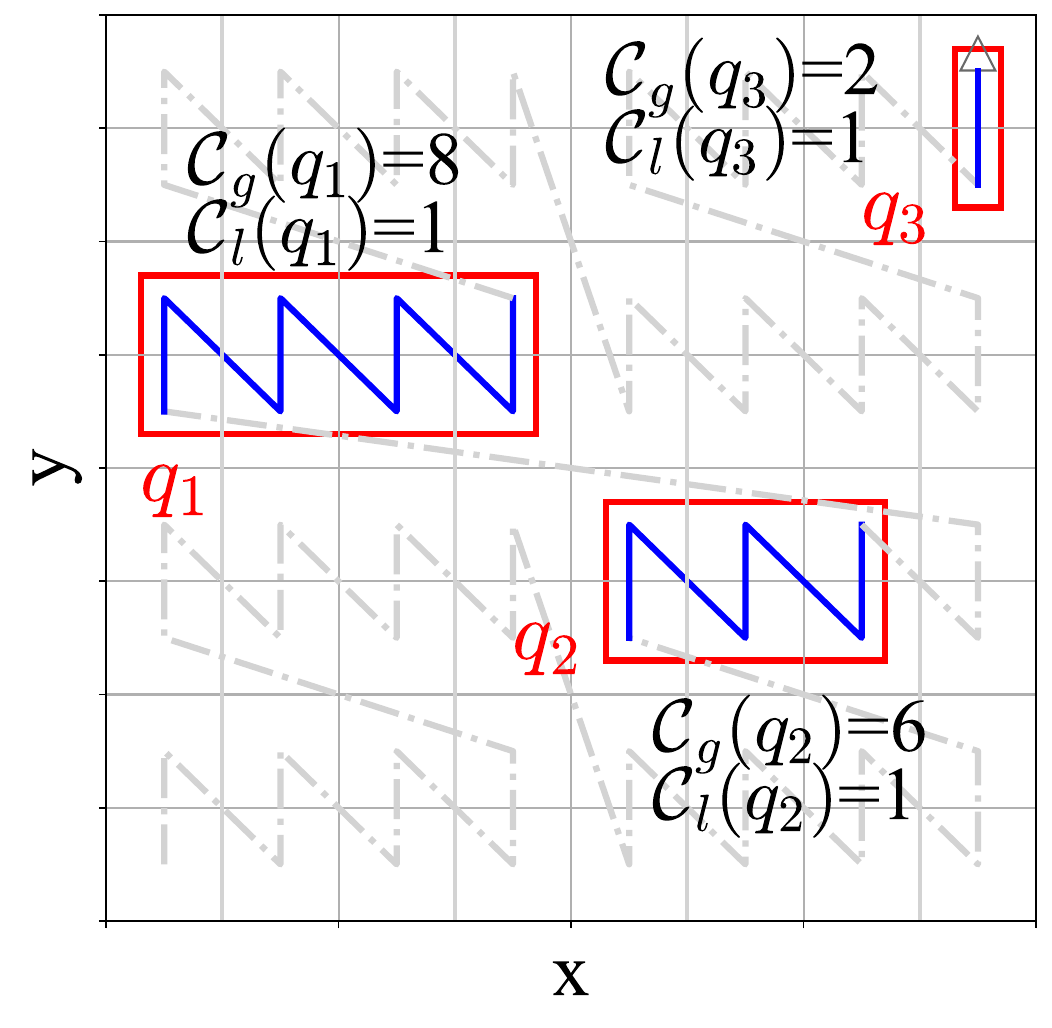}
	}
 \vspace{-3mm}\\
	\subfloat[Learning through \method~\label{fig:DRL}]{
		\includegraphics[width=0.24\textwidth]{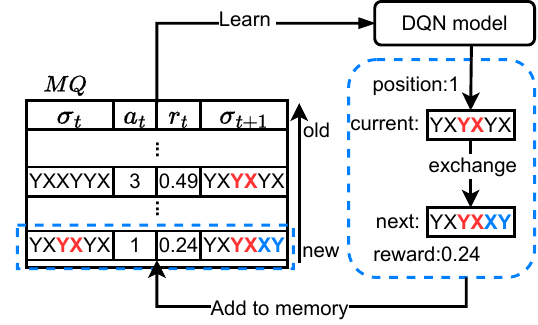}
	}
	\subfloat[Cost ratio vs. number of steps~\label{fig:RL_loss}]{
		\includegraphics[width=0.22\textwidth]{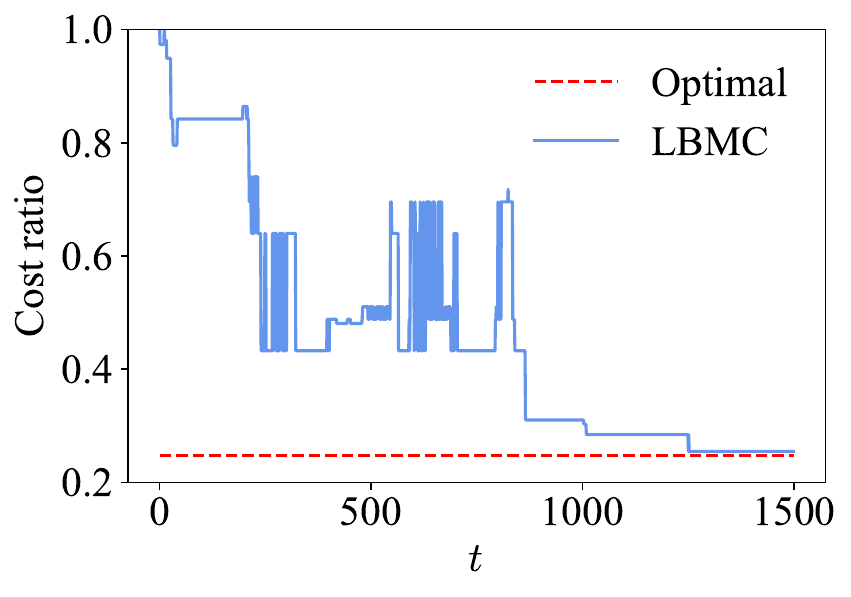}
	}	
	\caption{A BMC learning example.}\label{fig:RL_demo}
\end{figure}

\begin{example}
Figure~\ref{fig:RL_demo} illustrates \method\  with $\ell=3$ and  three queries $q_1$, $q_2$, and $q_3$. 
The initial BMC  $\sigma_1 = $ YX\uwave{X}\underline{Y}YX has an (estimated) query cost of $\mathcal{C}_1=175$ (Figure~\ref{fig:RL_0}).
We select position $a_1=3$ and swap the $3$rd and the $4$th bits to get $\sigma_2 = $ YX\underline{Y}\uwave{X}Y\underline{\underline{X}} such that the cost is decreased to $\mathcal{C}_2=90$ (Figure~\ref{fig:RL_1}). Next, we select position $a_2=1$ and swap the $1$st and the $2$nd bits to get $\sigma_3 = $ YXYX\underline{\underline{X}}Y with cost $\mathcal{C}_3=48$ (Figure~\ref{fig:RL_2}). 
We store all the intermediate results into memory $MQ$ for learning the DQN model in Figure~\ref{fig:DRL}, where we show the BMCs without encoding. 
Figure~\ref{fig:RL_loss} shows the cost ratios, i.e.,  
$\mathcal{C}_t/\mathcal{C}_1$, which decrease as $t$ increases (Figures~\ref{fig:RL_0} to~\ref{fig:RL_2} are three of the steps). The learned BMC approaches the optimum in this process. 
\end{example}

\textbf{Algorithm cost.} 
\method\ involves $T\cdot M$ iterations that each involves three key operations: bit-swap position prediction, reward calculation (cost estimation), and model training. Their costs are $O(1)$, $O({\mathcal{C}_t})$, and $O(\mathbb{T}_{\theta})$, respectively. 
The total time cost is then $O(T\cdot M\cdot (1+{\mathcal{C}_t}+\mathbb{T}_{\theta}))$. 
Here, $T\cdot M$ is a constant, while 
$O(\mathbb{T}_{\theta})$ is determined by the model structure. Our cost estimation results in ${O(\mathcal{C}_t}) = O(1)$, thus enabling an efficient  BMC search.

\section{Experiments}~\label{sec:experiments}
We aim to evaluate the (1)~efficiency and (2)~effectiveness of the proposed cost estimation algorithms, as well as (3)~\method\ vs. other SFCs, including the learning-based ones.

\subsection{Experimental Settings}~\label{sub:exp_set}
Our cost estimation algorithms (i.e., GC and LC) and BMC learning algorithm (i.e., \method) are implemented in Python (available at \url{https://anonymous.4open.science/r/LearnSFC-B6D8}). 
The learning of BMC is supported by TensorFlow.
We run experiments on a desktop computer 
running 64-bit Ubuntu 20.04 with a 3.60 GHz Intel i9 CPU,  64~GB RAM, and a 500 GB SSD.


\textbf{Datasets.} 
We use two real datasets: \textbf{OSM}~\cite{osm} and \textbf{NYC}~\cite{NYC}.
OSM contains 100 million 2-dimensional location points (2.2 GB). 
NYC contains some 150 million yellow taxi transactions (8.4 GB). After cleansing incomplete records, we retain the pick-up locations (2-dimensional points) of 100 million records. 
Additionally, we follow the study of the state-of-the-art competitor, the BMTree~\cite{BMTree}, and use two synthetic datasets, each with 100 million points: \textbf{UNI} and \textbf{SKEW}, which follow uniform and skewed distributions.

\textbf{Queries.} 
We again follow the BMTree study and generate synthetic query workloads.
Specifically, 1,000 synthetic queries are used for SFC learning, while 2,000 queries are generated separately for testing. The queries are of uniform size and follow the distributions of their respective datasets. To assess our cost estimation algorithms (Sections~\ref{subsection:efficiency} and~\ref{subsection:effectiveness}), we employ square queries, since the query shape does not impact the cost estimation time.

\textbf{Evaluation metrics.}
The core evaluation metrics used are (1)~the \textbf{cost estimation time}, (2) the \textbf{average number of block accesses per query} when using different SFC ordering for query processing (in PostgreSQL), and (3) the \textbf{SFC learning time}.



\textbf{Parameter settings.}
Table~\ref{tab:para} summarises the parameter values used, with default values in \textbf{bold}. In the table, $n$ denotes the number of queries; $\delta$ denotes the edge length of a query; $d$ denotes the data dimensionality; and $N$ denotes the dataset cardinality. We randomly sample from the datasets described above to obtain datasets of different cardinalities. 

For SFCs, a key parameter is the number of bits $\ell$, which impacts the curve value mapping efficiency substantially. To evaluate the cost estimation efficiency, we restrict $\ell$ to 18, beyond which a naive local cost baseline becomes computationally infeasible. In later experiments, we set $\ell=20$ following the BMTree to balance the computational costs of curve value mapping and cost estimation.

The BMTree has two additional parameters: 
the dataset sampling rate $\rho$ to form a subset for query cost estimation, and the depth $h$ of space partitioning.

\vspace{2mm}
\begin{table}[h]
        \centering
        \small
        \caption{Parameter settings.}
        \label{tab:para}
        \setlength{\tabcolsep}{1.0mm}
         \renewcommand{\arraystretch}{0.9} 
        \begin{tabular}{l|c|l}
            \toprule   
            Experiments & Parameter & Values \\
            \midrule
            \midrule
            \multirow{5}{*}{}Cost   & $n$ & $2^0$, $2^1$, $2^2$, $2^3$, \pmb{$2^4$}, $2^5$, $2^6$, $2^7$, $2^8$, $2^9$, $2^{10}$\\
            \cline{2-3}
            estimation
            &  $\delta(\times 2^{4})$  & \textbf{1}, 2, 4, 8, 16\\
            \cline{2-3}
            efficiency & $\ell$ & \textbf{10}, 12, 14, 16, 18\\
            \cline{2-3}
               & $d$ & \textbf{2}, 3, 4 \\
            \hline
            \multirow{4}{*}{}Cost & $N$ & $10^4$, $10^5$, $10^6$, \pmb{$10^7$}, $10^8$\\
            \cline{2-3}
            estimation
            &  $\rho (\times 10^{-3})$  & 0.1, 0.25, 0.5, 0.75, \textbf{1}, 2.5, 5, 7.5, 10\\
            \cline{2-3}
            effectiveness & $h$ & 5, 6, 7, 8, 9, \textbf{10}\\
            \cline{2-3}
               & $n$ & 100, 500, 1000, 1500, \textbf{2000} \\
            \cline{2-3}
               & datasets & \textbf{OSM}, SKEW \\
            \hline
           
            \multirow{4}{*}{} Query & 
            $N$ & $10^4$, $10^5$, $10^6$, \pmb{$10^7$}, $10^8$\\
            \cline{2-3}
             efficiency&aspect ratio  & $16:1$, $4:1$, $1:1$, $1:4$, \pmb{$1:16$}\\
            \cline{2-3}
             &$\delta (\times 2^{6})$  & 1, 2, \textbf{4}, 8, 16\\
             \cline{2-3}
               & datasets & \textbf{OSM}, NYC, UNI, SKEW, \\
            \bottomrule
        \end{tabular}
    \end{table}


\subsection{Cost Estimation Efficiency}~\label{subsection:efficiency}
We first evaluate the efficiency of our algorithms (excluding initialization) to compute the global cost \textbf{GC} and the local cost \textbf{LC} (Algorithm~\ref{alg:read_table}), which are based on Equations~\ref{equ:global_cost} and \ref{equ:connection_md}. 
We use \textbf{IGC} and \textbf{ILC} to denote the initialization steps of the two costs, respectively. As there are no existing efficient algorithms to compute these costs, we compare with baseline algorithms based on Equations~\ref{equ:global_cost_naive} and \ref{equ:local_cost}, denoted by \textbf{NGC} and \textbf{NLC}. 

We vary the number of queries $n$, the query size (via $\delta$), and the number of bits $\ell$. We run experiments for 2- to 4-dimensional spaces. Due to page limits, we focus on the 2-dimensional space (the algorithms' comparative results are similar for $d\in\{3,4\}$). As the cost estimation is data independent, a dataset is not needed to study their efficiency. The queries are generated at random locations.



\begin{figure}[h]
\centering
	\subfloat[Varying $n$~\label{fig:global_cost_num}]{
		\includegraphics[width=0.24\textwidth]{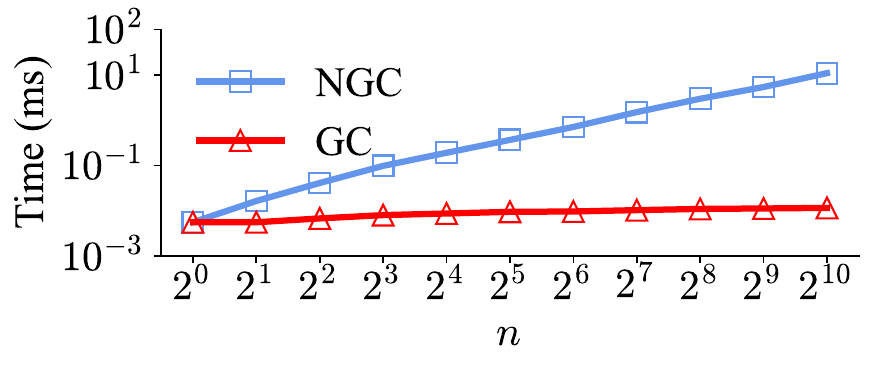}
        \hspace{-3mm}
	}
	\subfloat[Varying $\delta$~\label{fig:global_cost_size}]{
		\includegraphics[width=0.24\textwidth]{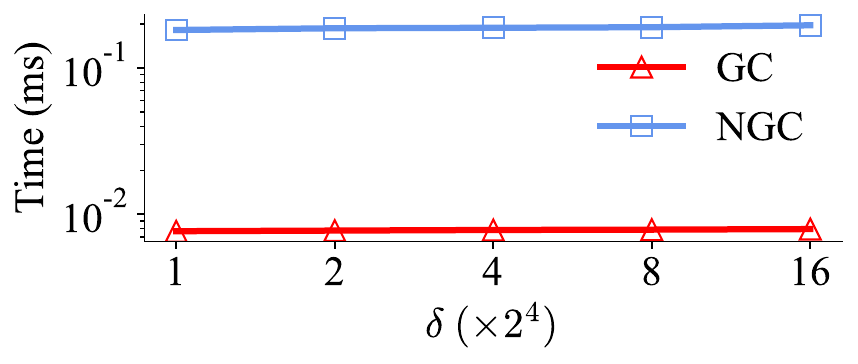}
	}\\
	\vspace{-5mm}
	\subfloat[Varying $\ell$ ~\label{fig:global_cost_bit}]{
		\includegraphics[width=0.24\textwidth]{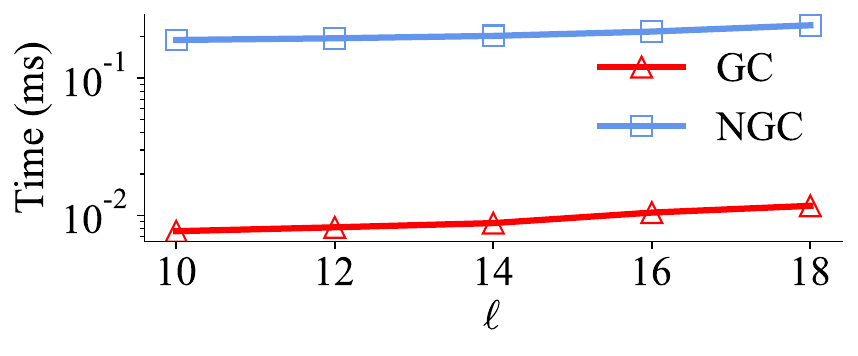}
        \hspace{-3mm}
	}
	\subfloat[Varying $d$~\label{fig:global_cost_dim}]{
		\includegraphics[width=0.24\textwidth]{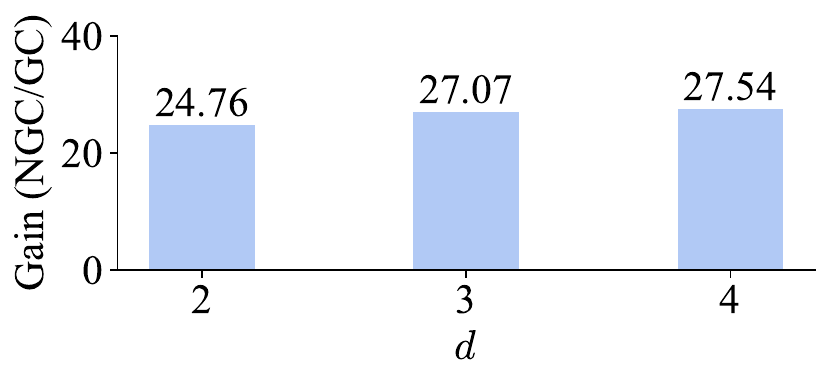}
	}
	\caption{Running times of global cost estimation.}\label{fig:global_cost_1}
\end{figure}

\subsubsection{Efficiency of GC}
Figures~\ref{fig:global_cost_num} and~\ref{fig:global_cost_size} show the impact of $n$ and~$\delta$, respectively. 
Since GC takes $O(d\cdot\ell)$ time to compute (after the initialization step), its running time is unaffected by $n$ and~$\delta$. NGC takes $O(n\cdot d\cdot\ell)$ time. Its running time grows linearly with $n$ and is unaffected by $\delta$ as shown in the figures.  
Figure~\ref{fig:global_cost_bit} shows that the running times of GC and NGC both increase with $\ell$, which is consistent with their time complexities. Since the relative performance of our algorithm and the baseline is stable when $\ell$ is varied, we use a default value of 10 instead of the maximum value 18 as mentioned earlier, to streamline this set of experiments. 
Figure~\ref{fig:global_cost_dim} shows the impact of $d$. Here, we show the performance gain (i.e., the running time of NGC over that of GC) instead of the absolute running times, which are of different scales when $d$ is varied such that it is difficult to observe the relative performance. We see that GC is faster than NGC by 24x.
Overall, GC is consistently faster than NGC, with up to more than an order of magnitude performance gain, which confirms the high efficiency of GC. 

\begin{figure}[h]
	\subfloat[Varying $n$ ~\label{fig:local_cost_num}]{
		\includegraphics[width=0.24\textwidth]{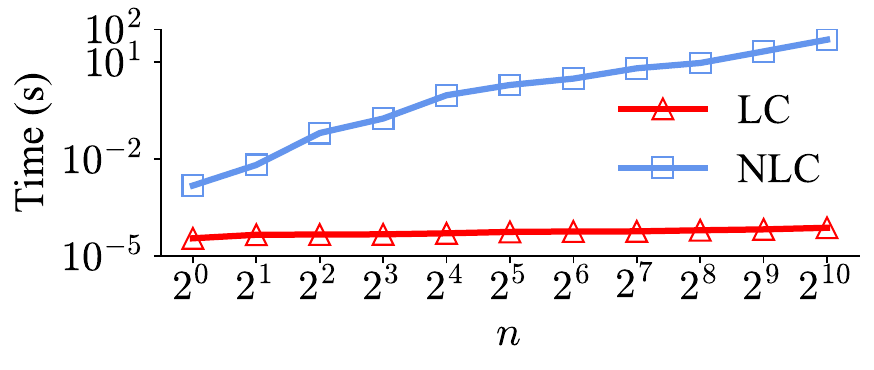}
        \hspace{-3mm}
	}
	\subfloat[Varying $\delta$ ~\label{fig:local_cost_size}]{
		\includegraphics[width=0.24\textwidth]{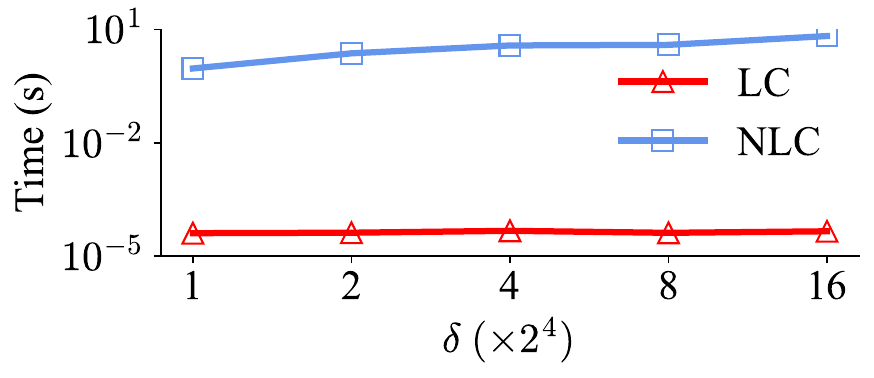}
	}\\
    \vspace{-5mm}
    \subfloat[Varying $\ell$~\label{fig:local_cost_bit}]{
		\includegraphics[width=0.24\textwidth]{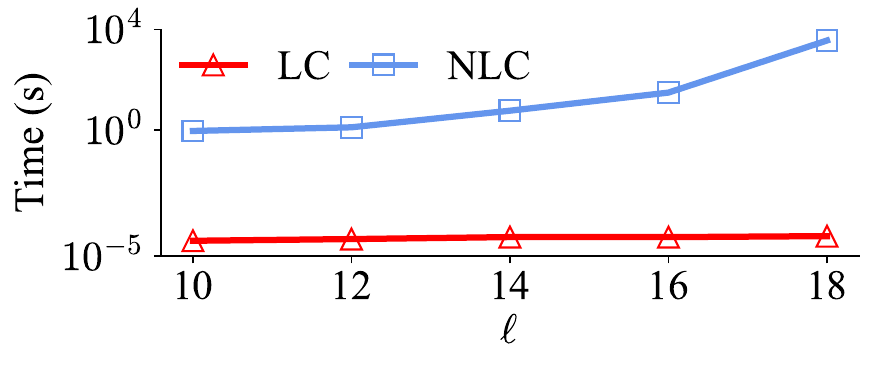}
            \hspace{-3mm}
	}
	\subfloat[Varying $d$ ~\label{fig:local_cost_dim}]{
		\includegraphics[width=0.24\textwidth]{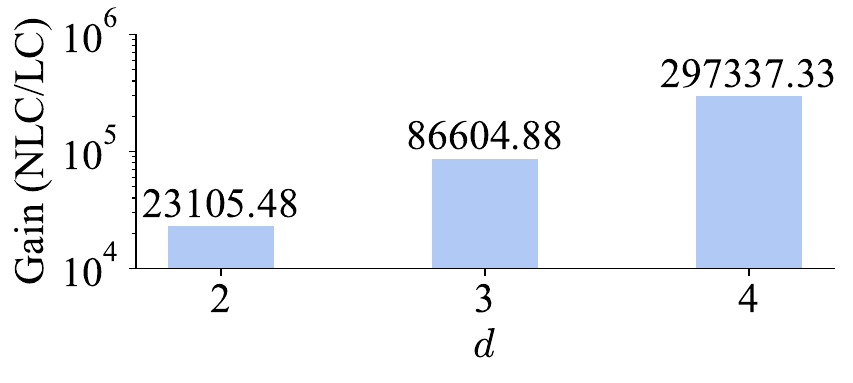}
	}
	\caption{Running times of local cost estimation.}\label{fig:local_cost_1}
\end{figure}

\subsubsection{Efficiency of LC}
Figures~\ref{fig:local_cost_num} to~\ref{fig:local_cost_dim} show the running times of computing local costs. The performance patterns of LC and NLC are similar to those observed above for GC and NGC, and they are consistent with the cost analysis in Section~\ref{subsec:local_cost}. 
The performance gains of LC are even larger, as its pre-computed pattern table enables extremely fast local-cost estimation. As Figure~\ref{fig:local_cost_dim} shows, LC outperforms NLC by five orders of magnitude when $d = 4$. 


\subsubsection{Initialization Costs of GC and LC}
Table~\ref{tab:init_vs_naive_global} shows the running times of IGC and ILC, which increase with $n$, because the initialization steps need to visit all range queries to compute a partial global cost and prepare the pattern tables, respectively. 
These running times are smaller than those of NGC and NLC, confirming the efficiency of the proposed cost estimation algorithms. Similar patterns are observed when varying $\delta$, $\ell$, and $d$, which are omitted for brevity. We do not report the result when $n=2^{0}$ (i.e., $n=1$) as no initialization is needed for a single query. 
\vspace{2mm}
\begin{table}[h]
    \centering
    \small
    \setlength{\tabcolsep}{2pt}
    \caption{Initialization costs of GC and LC (Varying $n$).}
    \label{tab:init_vs_naive_global}
    \begin{tabular}{c|c|c|c|c|c|c|c|c|c|c}
      \toprule  
      $n$ & $2^1$ & $2^2$ & $2^3$& $2^4$ &$2^5$ 
      & $2^6$& $2^7$& $2^8$& $2^9$& $2^{10}$ \\
      \midrule
      \midrule
      IGC (ms) & \textbf{0.03}& \textbf{0.05} & \textbf{0.08}& \textbf{0.15} &\textbf{0.27} & \textbf{0.52} & \textbf{1.06} & \textbf{1.93} &\textbf{4.07} & \textbf{7.79}\\
      \hline
      NGC (ms) & 0.03& 0.05 & 0.10& 0.18 & 0.36 & 0.70 & 1.50 & 2.96 & 5.37 & 10.86\\
      \hline
      \hline
      ILC (s) & \textbf{0.01}& \textbf{0.01}& \textbf{0.02} & \textbf{0.06}& \textbf{0.12} & \textbf{0.23} & \textbf{0.48} & \textbf{0.95} & \textbf{1.83} & \textbf{3.63} \\
      \hline
      NLC (s) & 0.01& 0.06 & 0.18 & 0.93 & 1.93 & 3.03 & 6.31 & 9.21 & 20.98 & 48.22\\
      \bottomrule
    \end{tabular}
\end{table}

\subsection{Effectiveness of Cost Estimation}~\label{subsection:effectiveness}
We next explore the applicability and effectiveness of our GC and LC cost estimations by using them to replace the built-in cost estimations of the state-of-the-art SFC learning algorithm, the 
BMTree. We denote the resulting variants by 
\textbf{BMTree-GC} and \textbf{BMTree-LC}. The original BMTree uses a data sampling-based empirical cost estimation method. We denote it as \textbf{BMTree-SP}.

We report the time cost of reward calculation for the three variants, as the other steps of the variants are the same. After the SFCs are learned by the three variants, we build a B$^+$-tree with each SFC in PostgreSQL to index the input dataset. We measure the average number of block accesses as reported by PostgreSQL to process each of the queries as described earlier. 

\subsubsection{Varying the Dataset Cardinality}~\label{subsubsec:vary_cardinality}
\begin{figure}[h]
\centering
	\subfloat[Reward calculation time~\label{fig:vary_datasize_time}]{
		\includegraphics[width=0.24\textwidth]{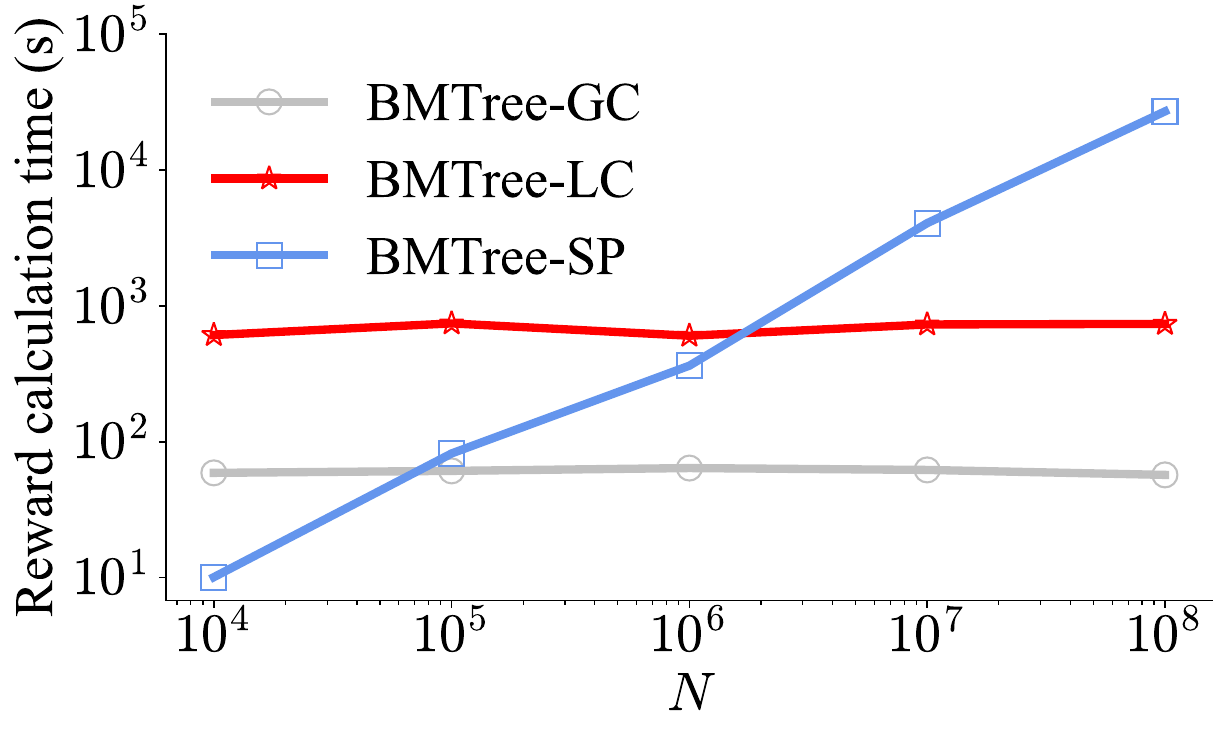}
        \hspace{-3mm}
	}
	\subfloat[Query processing cost~\label{fig:vary_datasize_query}]{
		\includegraphics[width=0.24\textwidth]{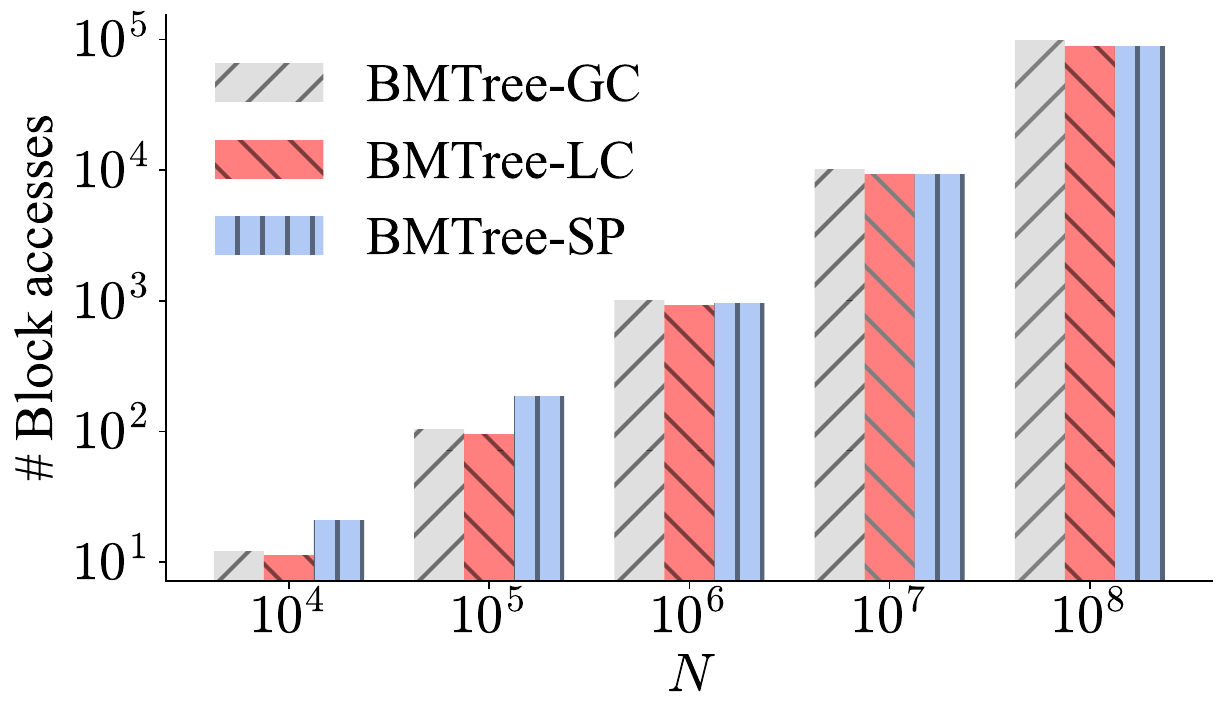}
	}
	\caption{Varying the dataset cardinality (OSM).}\label{fig:vary_datasize}
\end{figure}

We start by varying the dataset cardinality $N$ from $10^4$ to $10^8$. 
Figure~\ref{fig:vary_datasize} shows the results on the OSM dataset (the results on the other datasets show similar patterns and are omitted for brevity; same below).
BMTree-GC and BMTree-LC have constant reward calculation times, since GC and LC are computed in constant times. In comparison, the reward calculation time of BMTree-SP increases linearly with the dataset cardinality, as BMTree-SP builds intermediate index structures based on sampled data points for query cost estimation. When $N$ increases, the number of sampled data points also increases. At $N = 10^8$ (the default sampling rate is $\rho = 0.001$, i.e., BMTree-SP is run on a sampled set of $10^5$ points), the reward calculation time of BMTree-SP (more than 7 hours) is 36x and 474x higher than those of BMTree-LC (737 s) and BMTree-GC (57 s).

In terms of the query costs, the indices built using all three algorithms require more block accesses as $N$ increases, which is expected. Importantly, all three algorithms incur similar numbers of block accesses given the same $N$ value. This suggests that the GC and LC cost estimations can be applied to improve the curve learning efficiency of the BMTree without adverse effects on the query efficiency. In general, BMTree-LC offers lower query costs than BMTree-GC. Thus, applications that are more sensitive to query costs may use BMTree-LC, while those that are more sensitive to index building costs may use BMTree-GC.

\subsubsection{Varying the Number of Queries}
\begin{figure}[h]
\centering
	\subfloat[Reward calculation time~\label{fig:vary_query_num_time}]{
		\includegraphics[width=0.24\textwidth]{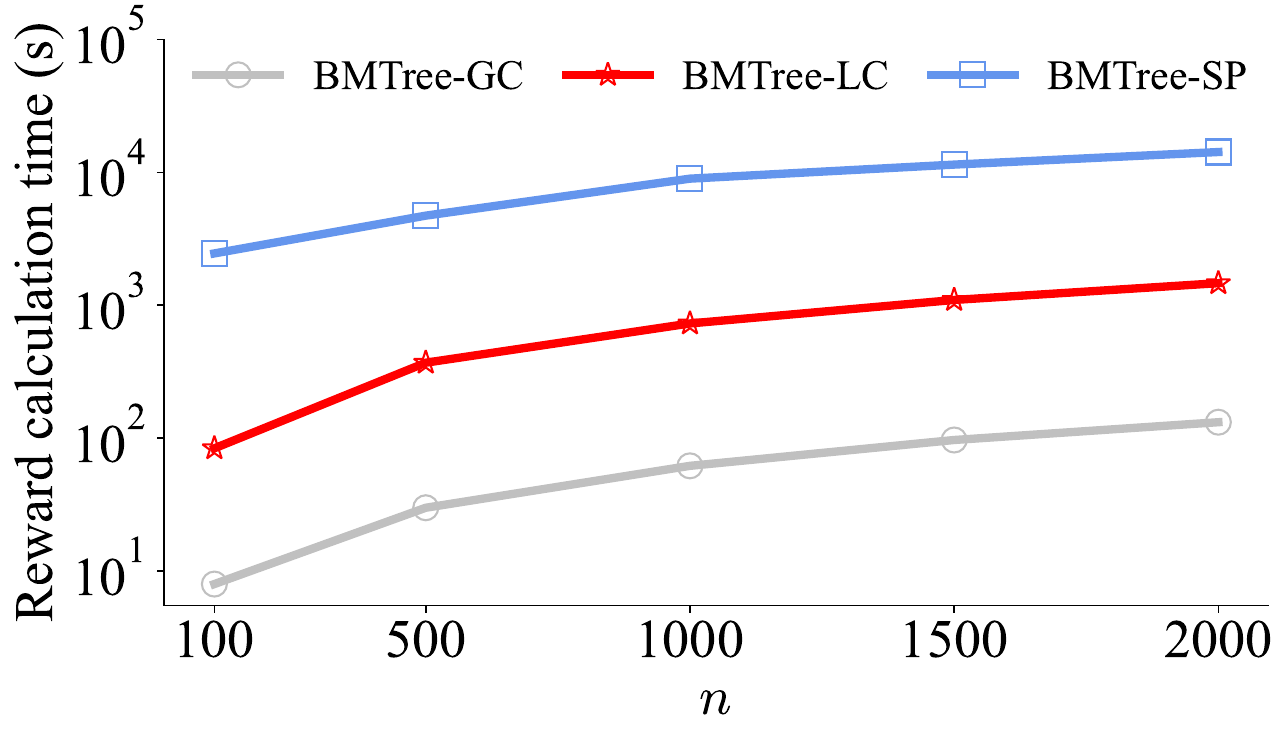}
        \hspace{-3mm}
	}
	\subfloat[Query processing cost~\label{fig:vary_query_num_query}]{
		\includegraphics[width=0.24\textwidth]{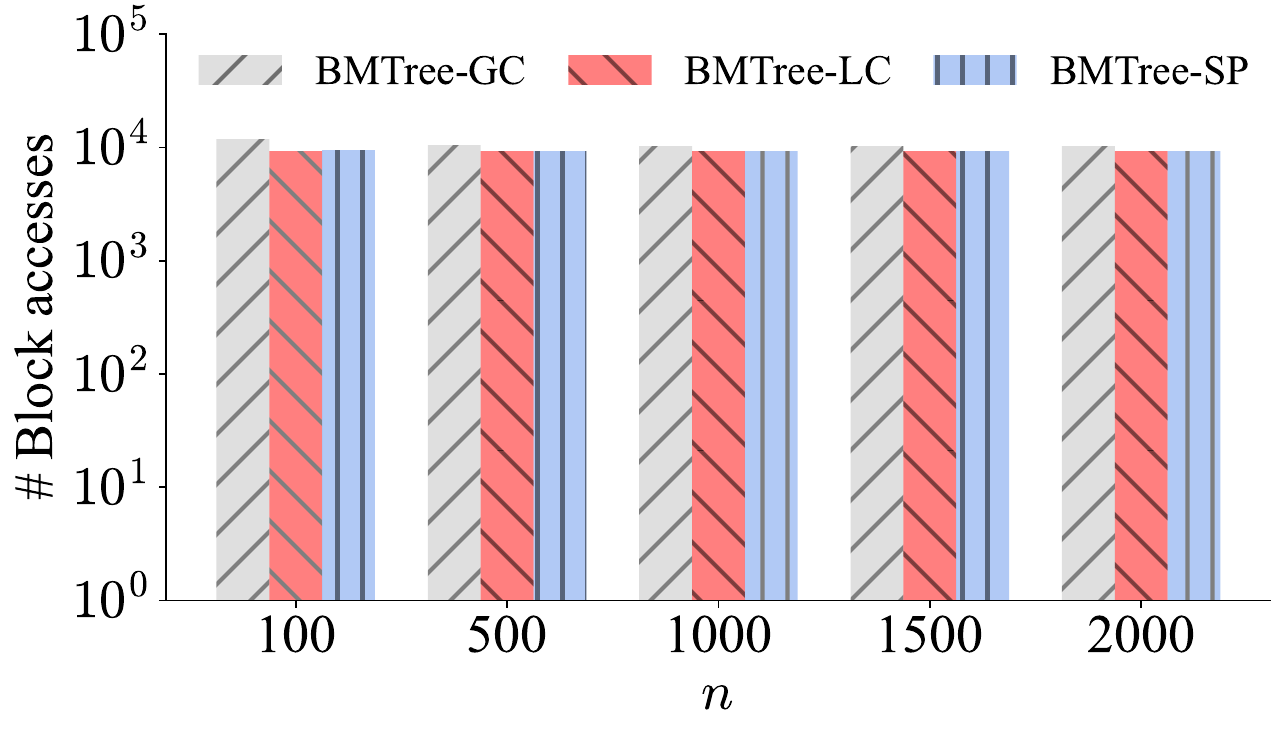}
	}
	\caption{Varying the number of queries (OSM).}\label{fig:vary_query_num}
\end{figure}

Next, we vary the number of queries used in curve learning, $n$, from 100 to 2,000. We see that BMTree-LC and BMTree-GC consistently outperform BMTree-SP by one and two orders of magnitude in terms of the reward calculation time, respectively (Figure~\ref{fig:vary_query_num_time}). We note that, now the computation times of BMTree-LC and BMTree-GC vary with $n$, which differs from what was reported in Figures~\ref{fig:global_cost_num} and~\ref{fig:local_cost_num}. This happens because the BMTree uses different BMCs in different sub-spaces to accommodate different data and query patterns. As there are more queries, more different patterns may need to be considered, resulting in more different BMCs, each of which requires a different GC and LC cost estimation. Thus, the cost estimation costs grow with the number of queries $n$. 

Meanwhile, the query costs of the three algorithms are again close, e.g., 9,199, 9,248, and 10,462, for BMTree-LC, BMTree-SP, and BMTree-GC, respectively, when $n$ is 1,500. The higher query cost of BMTree-GC shows that while GC is extremely simple and efficient, it may not find the most query-efficient curves, which underlines the importance of the LC cost estimation algorithm. 

We further observe a slight drop in the number of block accesses as $n$ increases. Intuitively, using more queries for curve learning can lead to curves that better suit the query workload. 



\subsubsection{Varying the Sampling Rate and the Depth of the BMTree}
Two alternative approaches to improve the curve learning efficiency of the BMTree are (1) to reduce its data sampling rate $\rho$ and (2) to reduce the depth of its space partitioning $h$. 

In this set of experiments, we study how these two parameters impact the reward calculation time and the query cost of the resulting SFCs. In particular, we vary $\rho$ from $10^{-4}$ to $10^{-2}$ (a total of 9 values, cf.~Table~\ref{tab:para}), and we vary  $h$ from 5 to 10. 

\begin{figure}[h]
\centering
	\subfloat[SKEW ~\label{fig:vary_depth_sample_rate_SKEW}]{
		\includegraphics[width=0.24\textwidth]{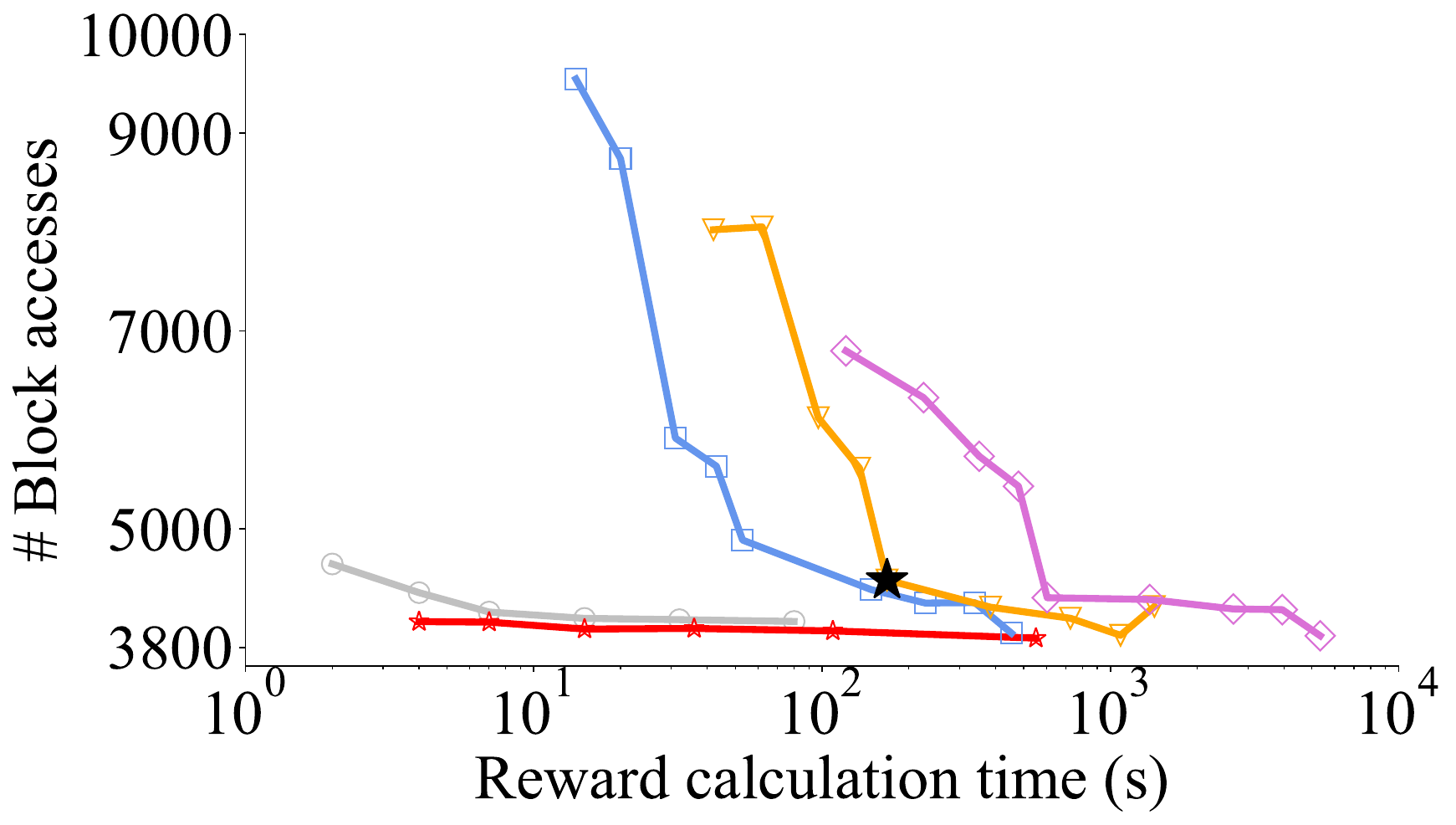}}
        \subfloat[OSM ~\label{fig:vary_depth_sample_rate_OSM}]{
	\hspace{-2mm}	\includegraphics[width=0.24\textwidth]{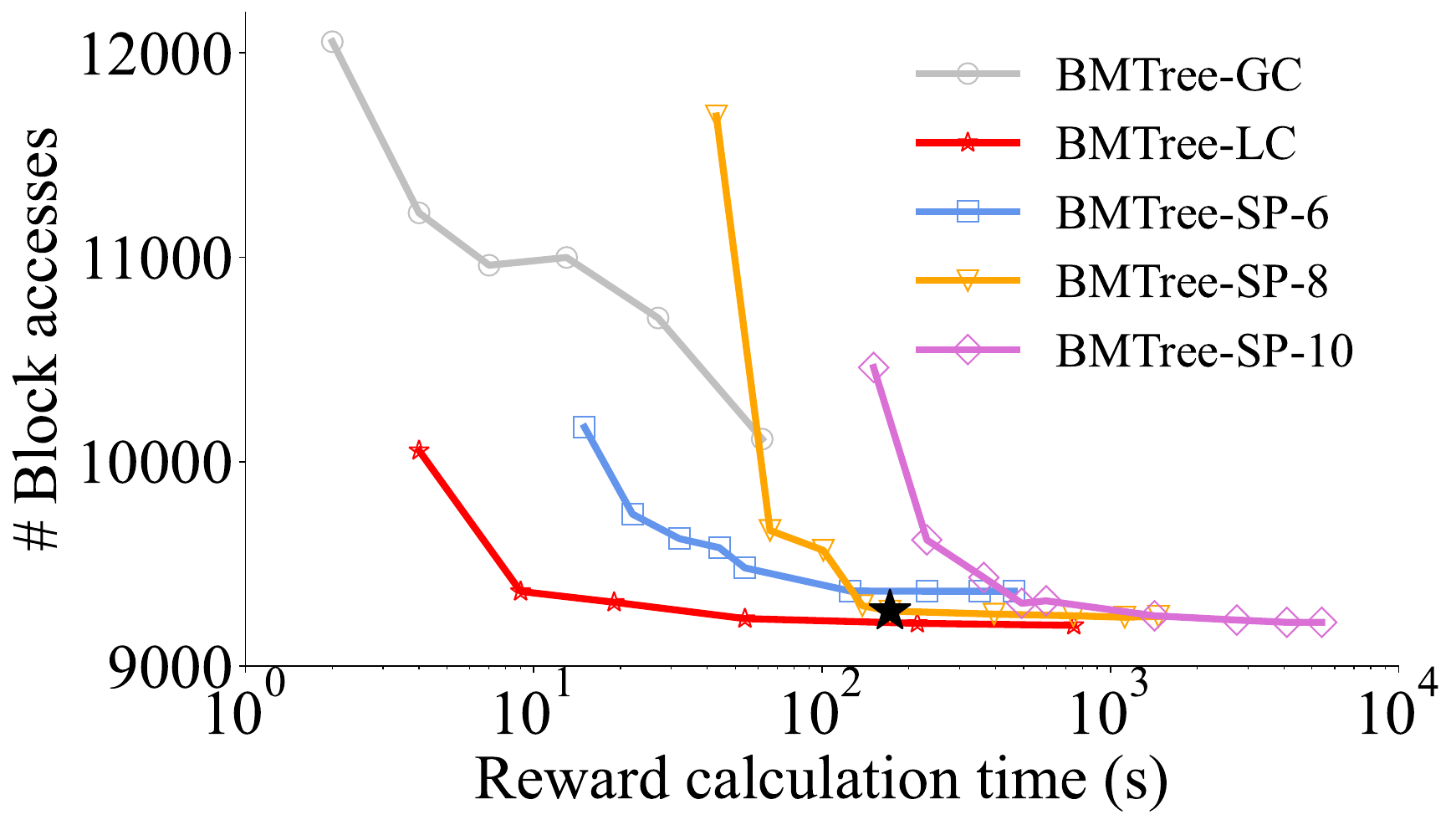}
	}
	\caption{Varying the sampling rate and the space partitioning depth of the  BMTree. }\label{fig:vary_depth_sample_rate}
\end{figure}

Figure~\ref{fig:vary_depth_sample_rate} plots the results on the SKEW and OSM datasets. BMTree-SP has three result polylines: BMTree-SP-6, BMTree-SP-8, and BMTree-SP-10, each of which uses a different $h$ value, while the points on each polyline represent the results of different $\rho$ values (points on the right come from larger $\rho$ values).

BMTree-GL and BMTree-LC are plotted with one polyline each, as they are not impacted by $\rho$. The points on these polylines represent the results of different values of $h$ (points on the right correspond to larger $h$ values). 

We see that a larger $h$ value tends to lead to lower query costs, while it also yields a longer reward calculation time. 
Powered by the LC cost estimation algorithm, BMTree-LC reduces the reward calculation time by at least an order of magnitude while achieving the same level of query costs (i.e., its curve lies at the bottom left of the figure). BMTree-GC can also be very fast at reward calculation, while it may suffer at query performance.

\subsection{Query Efficiency with BMC Learning}~\label{subsec:exp_curve}
We proceed to study the BMC learning efficiency of \textbf{\method} and the query efficiency of the indices built using the learned BMCs. 

\textbf{Competitors.} 
We compare with five different SFC-based ordering techniques. 
(1)~\textbf{QUILTS}~\cite{QUILTS} orders data points by a BMC derived by a curve design method as described in Section~\ref{sec:related_work}. We implement it according to its paper as the source code is unavailable. (2) \textbf{ZC}~\cite{ZRtree} orders data points by their Z-curve values. (3) \textbf{HC}~\cite{HRtree} orders data points by their Hilbert curve values. (4) \textbf{LC}, which is also called the C-Curve, orders data points lexicographically by their dimension values~\cite{QUILTS,BMTree}. 
(5) \textbf{BMTree}~\cite{BMTree} orders data points by multiple BMCs in different sub-spaces. We use its released code (with $h=8$ and $\rho = 0.001$ to balance the reward calculation time and the query costs, cf. the `$\star$'-points  on BMTree-SP-8 in Figure~\ref{fig:vary_depth_sample_rate}). We cannot compare with the recent learned SFC, LMSFC~\cite{LMSFC}, because its source code and some implementation details are unavailable. We do not compare with RSMI~\cite{RSMI} as it has been shown to be outperformed by the BMTree~\cite{BMTree}.

For all techniques, we use the curves obtained to order the data points and build B$^+$-trees in PostgreSQL for query processing, and we report the average number of block accesses as before.

\subsubsection{Overall Results}~\label{subsubsec:all_datasets}
\begin{figure}[h]
\centering
	\includegraphics[width=0.42\textwidth]{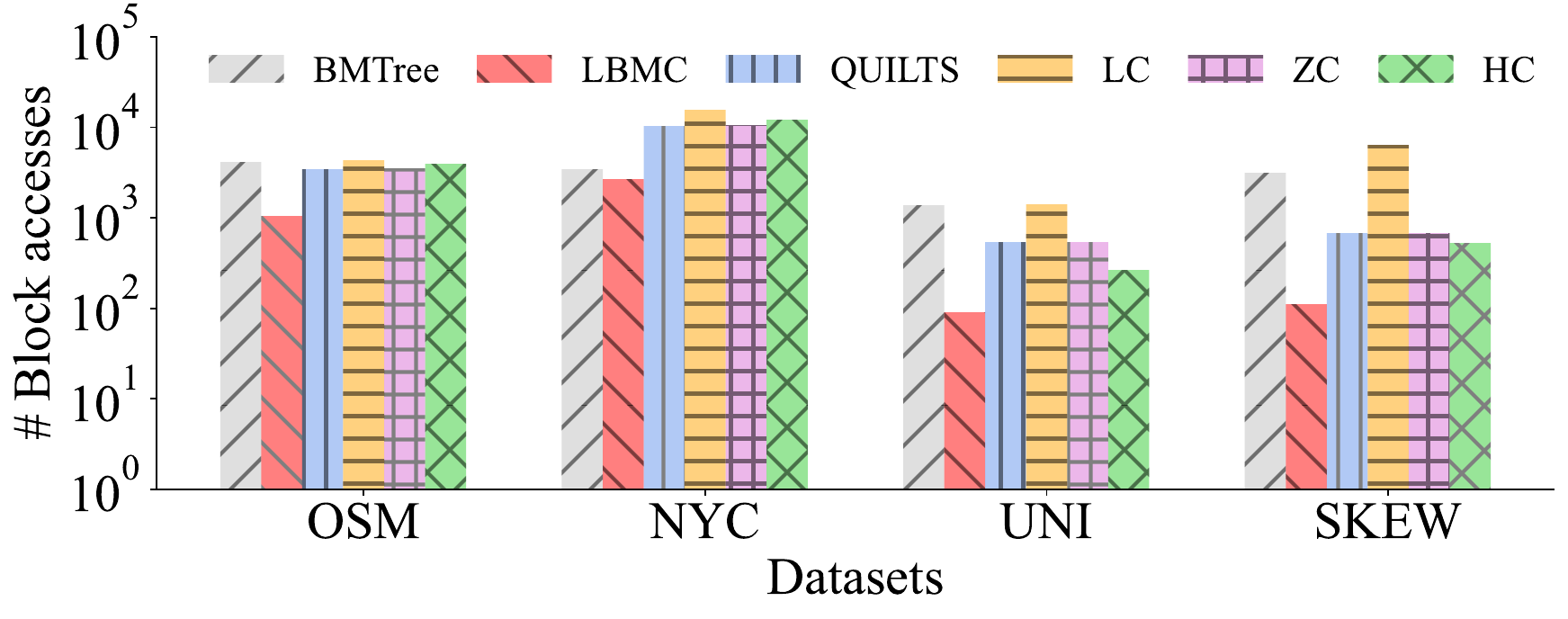}
	\caption{Block access over all datasets.}\label{fig:varying_dataset_block_access}
\end{figure}

Figure~\ref{fig:varying_dataset_block_access} shows the average number of block accesses on all four datasets. 
\method\ outperforms all competitors consistently.
On SKEW, the advantage of \method\ over the BMTree is the most pronounced.
It reduces the average number of block accesses by 28x (111 vs. 3,084) and by 6x (111 vs. 674) in comparison with the BMTree and QUILTS, respectively. 
On NYC, the advantage of \method\ over the BMTree is the least, yet it still requires only 2,638 block accesses which is fewer than that of the BMTree at 3,448.
These results suggest that \method\ is highly efficient at reducing the query costs across diverse datasets.

LC is the worst, which is expected as LC curves fail to preserve the data locality. The BMTree and QUILTS outperform LC, ZC, and HC on real data such as NYC, where they benefit more from the query based optimizations. However, there are no consistent results across the different datasets. We conjecture that fine-tuning of the parameter values of $h$ and $\rho$ may be needed for the BMTree over each different dataset. Such fine-tuning is not required by~\method.   

\vspace{1mm}

\subsubsection{Varying the Dataset Cardinality}~\label{subsubsec:varying_cardinality}
\begin{figure}[h]
\centering
	\includegraphics[width=0.42\textwidth]{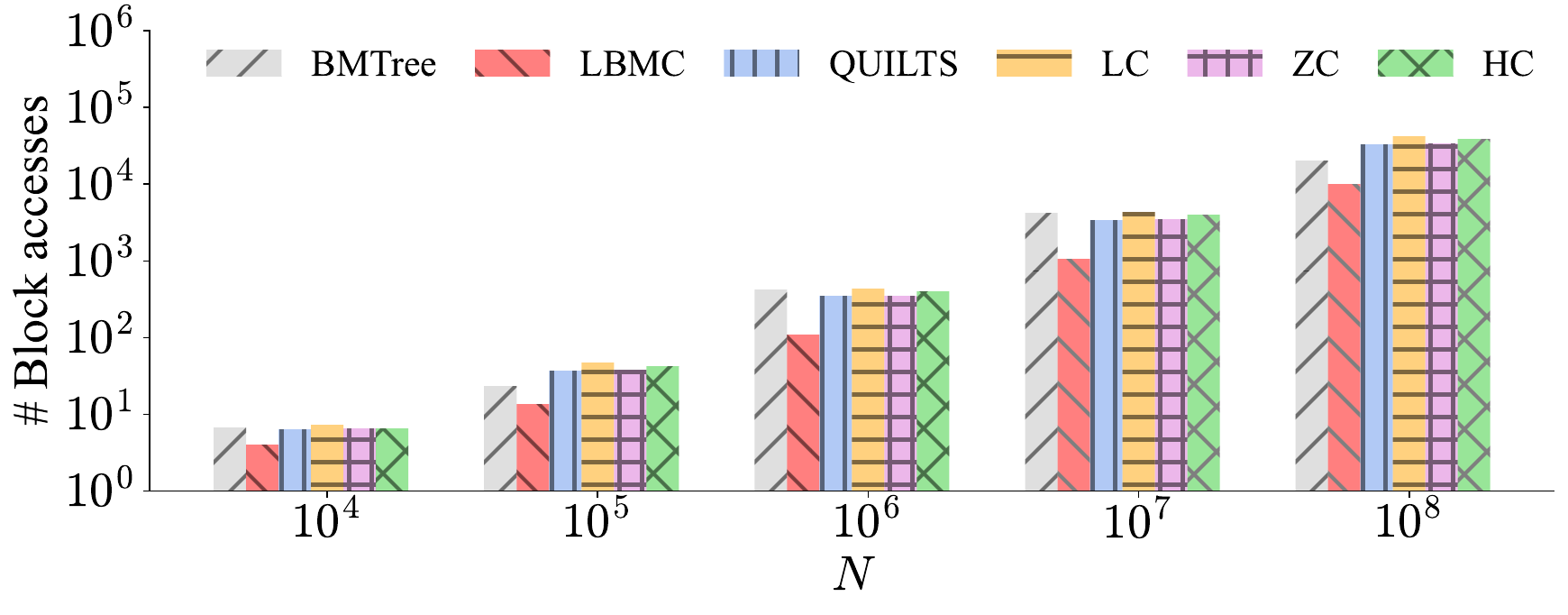}
	\caption{Varying cardinality (OSM).}\label{fig:vary_cardinality_block_access}
\end{figure}

We further study the impact of dataset cardinality $N$.  
Figure~\ref{fig:vary_cardinality_block_access} shows the results. Like before, the 
average number of block accesses increases with $N$, which is expected. 
\method is again the most efficient in terms of query costs, needing at least 39\% fewer block accesses than the BMTree (4.0 vs. 6.6 when $N = 10^4$), and the advantage is up to 74\%  (1,044 vs. 4,131 when $N = 10^7$).


\vspace{1mm}
\vspace{2mm}

We report the SFC learning times of the BMTree and \method\ when varying $N$ in Table~\ref{tab:extra_cost}. 
We see that \method\ is much faster than the BMTree at SFC learning and that the advantage grows with $N$. This is because the cost estimation (i.e., reward calculation) in the BMTree is much slower than that in \method, as shown in the last subsection. The cost estimation time dominates when there are more data points for the BMTree, while the cost estimation time of \method\ remains constant when varying $N$. 

LC, ZC, and HC are not learned, and they do not take any learning time. QUILTS takes less than 1 second, as it only considers a few curve candidates (which are generated based on query shapes) using a cost model. We have used our cost estimation algorithms in our implementation of QUILTS, as the original cost model is prohibitively expensive.


\vspace{2mm}
\begin{table}[h]
    \centering
    \small
    \caption{SFC learning time (seconds).}
    \label{tab:extra_cost}
    \begin{tabular}{c|r|r|r|r|r}
      \toprule  
      $N$ & $10^4$ & $10^5$ & $10^6$& $10^7$ &$10^8$ \\
      \midrule
      \midrule
      BMTree & 54 & 55 &  61 & 99 & 551 \\
      \hline
      \method\ & 15& 15 & 15& 15 & 15 \\
      \hline
      QUILTS (with our cost estimation) & 0.2& 0.2 & 0.2& 0.2 & 0.2 \\
      \bottomrule
    \end{tabular}
\end{table}


\subsubsection{Varying the Aspect Ratio of Queries}~\label{subsubsec:varying_aspect_ratio}
\begin{figure}[h]
\centering
	\includegraphics[width=0.42\textwidth]{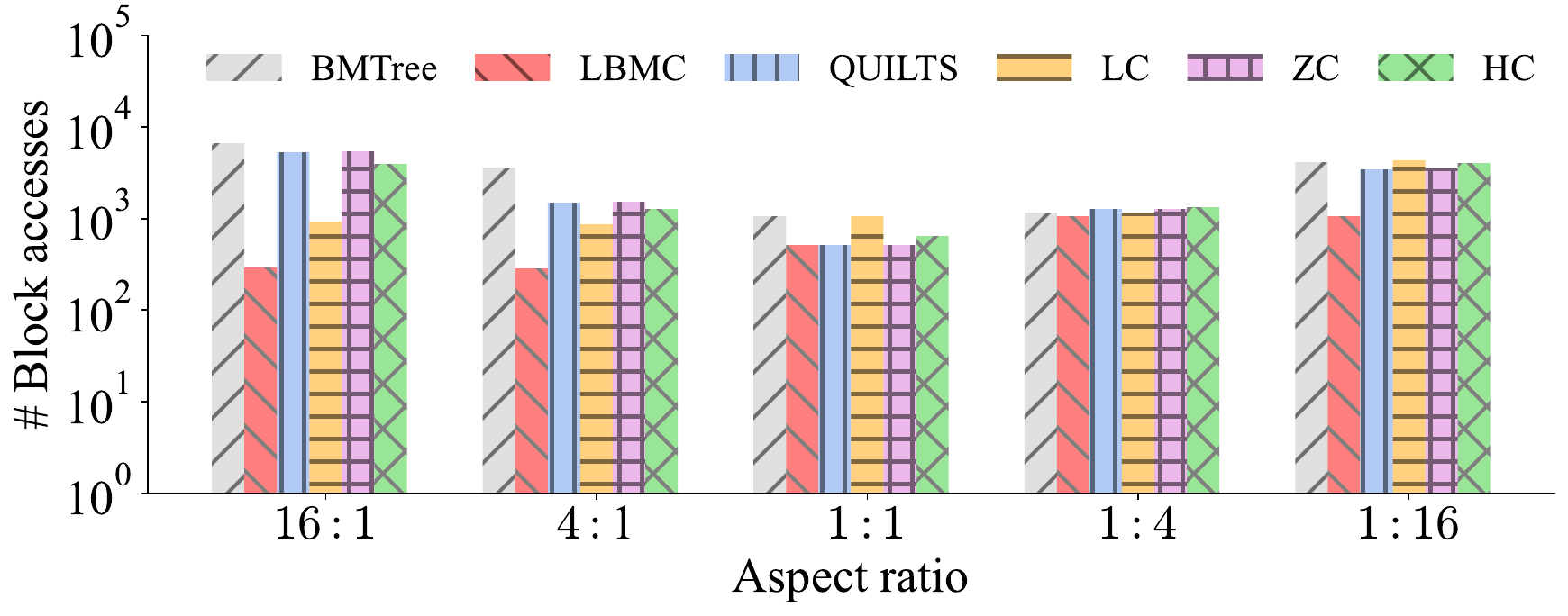}
	\caption{Varying the query aspect ratio (OSM).}\label{fig:varying_skewness_block_access}
\end{figure}

Figure~\ref{fig:varying_skewness_block_access} shows the query costs when varying the query aspect ratio. 
Here, \method\ shows a stronger advantage over the competitors on queries that are ``stretched'', while LC also better suits the queries that are long and thin (16:1) which is intuitive.  
When the aspect ratio is $1:1$, \method, QUILTS, and ZC share almost the same query performance because they all tend to form a `{\Large{\backwardsz}}' shape to fit square queries. 
The BMTree is again outperformed by \method, because of its less flexible learning scheme (i.e., learning for only up to $h$ bits), while \method\ can learn a BMC scheme with all $\ell$ bits ($\ell = 20$ by default). 


\subsubsection{Varying the Edge Length of Queries}~\label{subsubsec:varying_area}
\begin{figure}[h]
\centering
	\includegraphics[width=0.42\textwidth]{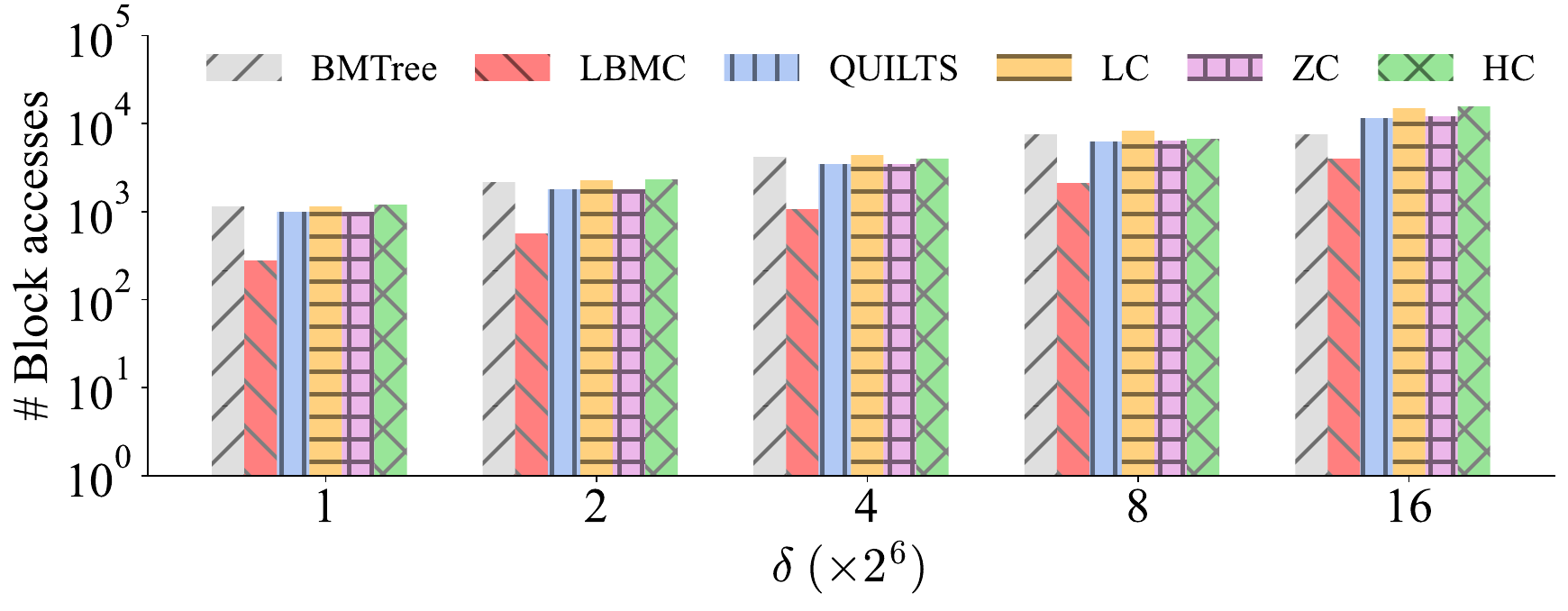}
	\caption{Varying the query edge  length (OSM).}\label{fig:varying_selectivity_block_access}
\end{figure}

Figure~\ref{fig:varying_selectivity_block_access}
shows that the average number of block accesses grows with the query edge length, as expected. 
Here, \method\ again outperforms the competitors consistently, further showing the robustness of \method.

\section{Conclusions and Future Work}~\label{sec:conclusions}
We studied efficient cost estimation for a family of SFCs, i.e., the BMCs. Our cost algorithms can compute the global and the local query costs of BMCs in constant time given $n$ queries and after an $O(n)$-time initialization. 
We extended these algorithms to the state-of-the-art curve learning algorithm, the BMTree, which originally measured the effectiveness of SFCs by querying the data points to be indexed. Experimental results show that the proposed algorithms are capable of reducing the cost estimation time of the BMTree by over an order of magnitude with little or no impact on the query efficiency of the learned curves. 

We further proposed a reinforcement learning-based curve learning algorithm. The result learned BMCs are shown to achieve lower query costs than those of the BMTree and other baselines under nearly all settings tested.

In future work, it is of interest to design cost estimation algorithms for non-BMCs, e.g., HC, and use learning-based techniques to build more efficient multi-dimensional indices.


\balance
\bibliographystyle{ACM-Reference-Format}
\bibliography{reference}

\end{document}